\documentclass{aa}

\usepackage{graphicx}
\usepackage{longtable}
\usepackage{supertabular}
\usepackage{natbib}
\usepackage{txfonts}
\begin{document}
  \title{Chemical abundances and kinematics of a sample of metal-rich 
barium stars
\thanks{Based on observations made with the 1.52m and 2.2m telescope 
at the European Southern Observatory (La Silla, Chile).}$^,$\thanks{Tables 2 and 4 are only available in electronic form
at the CDS via anonymous ftp to cdsarc.u-strasbg.fr (130.79.128.5)
or via http://cdsweb.u-strasbg.fr/cgi-bin/qcat?J/A+A/.}}

 \author{C.B. Pereira\inst{1}, J.V. Sales Silva\inst{1}, C. Chavero\inst{1},
   F. Roig\inst{1} \& E. Jilinski\inst{1,2}}

   \offprints{C.B. Pereira}

\institute{Observat\'orio Nacional, Rua Jos\'e Cristino, 77. 
CEP 20921-400. S\~ao Crist\'ov\~ao. Rio de Janeiro-RJ. Brazil.\\
    \and
Instituto de F\'{\i}sica, Universidade do Estado do Rio de Janeiro, 
Rua S\~ao Francisco Xavier 524, Maracan\~a, 200550-900 Rio de Janeiro-RJ, Brazil\\
   \email{claudio,joaovictor,carolina,froig,jilinski@on.br}}

  \date{Received ; accepted }  

\abstract 
{} 
{We determined the atmospheric parameters and abundance pattern for a
sample of metal-rich barium stars.}  
{We used high-resolution optical spectroscopy. Atmospheric parameters 
and abundances were  determined using the local thermodynamic equilibrium 
atmosphere models of Kurucz and the spectral analysis code MOOG.} 
{We show that the stars have enhancement factors, [s/Fe], from 0.25 to
1.16. Their abundance pattern of the Na, Al, $\alpha$-elements, and iron group 
elements as well as their kinematical properties 
are  similar to the characteristics of the other metal-rich and super 
metal-rich stars already  analyzed. We conclude that metal-rich barium stars 
do not belong to the bulge population. We also show that metal-rich barium
stars are useful targets for probing the s-process enrichment in 
high-metallicity environments.}  
{}


   \authorrunning{Pereira}
   \titlerunning{Metal rich barium stars}

   \maketitle

\section{Introduction}

\par Barium stars are chemically peculiar giant
stars displaying both anomalously high heavy-element (Z $>$ 30) abundances and
an overabundance of carbon compared to the Sun, with [C/Fe] varying from +0.4 to
+1.2 \citep{antipova2004,allenbarbuy2006,drakepereira2008,pereiradrake2009}.
In the mass-transfer hypothesis, the observed chemical peculiarities of these
stars are a consequence of mass transfer through stellar winds or through Roche-lobe
 overflow in a binary system from an AGB star (now the white dwarf) to a
less evolved companion, a barium giant or a subgiant CH star.  Therefore, the
study of their chemical peculiarities is important for  understanding 
the physics of accretion phenomena in chemically peculiar binary systems.

\par Barium giants are the largest
known sample among the chemically peculiar stars where the effects of the
s-process nucleosynthesis can be widely investigated and measured.
\cite{busso2001} use the [hs/ls]
\footnote{[hs/ls]$=\log{\rm(hs/ls)}_{\star}-\log{\rm (hs/ls)}_{\odot}$ where
  [hs] and [ls] are the mean abundances of the s-elements at the Ba and Zr
  peaks, respectively.} index as a function of metallicity as a criterion to
probe whether the current models of s-process nucleosynthesis in AGB stars
account for the observed abundance distributions.

\par Barium stars are found in the disk and in the halo of the Galaxy
 according to \cite{gomez1997} and \cite{mennessier1997}.
These two studies show that barium stars
are an inhomogeneous group that can also be divided into groups according to
their luminosities, kinematical and spatial parameters ($U_{0}$, $V_{0}$,
$W_{0}$ velocities and dispersion velocities and scale heights).  If the
barium star phenomenon is seen both in the disk and in the  halo,
they are very useful objects to probe the s-process
nucleosynthesis at different metallicities, in  different
population.  Barium stars have already been analyzed in the halo, such as HD 206983
\citep{junqueirapereira2001,drakepereira2008}, HD 10613
\citep{pereiradrake2009}, and HD 123396 \citep{allenbarbuy2006}, and their
connection with the metal-poor halo yellow symbiotic stars have already been
investigated \citep{jorissen2005,pereiradrake2009}.  Therefore, the
quantitatively confirmation of the overabundances of the s-process elements in
a sample of barium stars would  help to better constrain  the number
  of known barium stars, which it would be useful to compare with their
theoretical birthrate \citep{han1995}.  Motivated by the questions mentioned
above, we started a high-resolution spectroscopic survey of the barium stars
from the samples of \cite{macConnell1972} and \cite{bidelman1981} as well as
some stars from \cite{gomez1997}.  As a first result of 230  surveyed stars, we
have already discovered a new CH subgiant, BD-03$^\circ$3668
\citep{pereiradrake2011}.

\par In this paper we report another result from our 
survey, that is, the discovery of a small sample of 12 metal-rich
([Fe/H]$\geq$+0.1) barium stars. According to their metallicities, metal-rich
stars can be divided into two sub-samples \citep{grenon1972}. Stars with
metallicities 0.08 $<$ [Fe/H] $<$ +0.2 are the metal-rich ones and those with
metallicities +0.2 $<$ [Fe/H] $<$ +0.5 are the super metal-rich. According to
this scheme, we found 7 metal-rich and 5 super metal-rich barium stars.  The
``very metal-rich'' stars were first identified by \cite{arp1965}
investigating the nucleus of the galaxy NGC 6522.  \cite{arp1965} also derived
 their age to be between 10$^{9}$ to 10$^{10}$ Gyr.  \cite{spinrad1969}
surveyed a large group of K giants with metal abundances greater than that of
the Hyades and with abundance variations among their sample. These authors also found
the giants ages to be older than the disk star, \cite{grenon1999} also showed
that metal-rich stars have an age mixture ranging from 0.7 Gyr for the Hyades
generation to an intermediate generation of 3-4 Gyr and to an oldest
generation, 10 Gyr.  In the investigation of the photometric and
kinematic properties of a large sample of G and K giants, \cite{eggen1993}
found that 10\% of the stars in the solar neighborhood have metallicities
higher than +0.15 dex.  \cite{eggen1993} also identified a sample of barium
stars among the evolved stars in the old disk population that could be
metal-rich as well. However, none of the stars presented in Table 11 of his
paper can  be classified as metal-rich or super metal rich (HD 46407,
[Fe/H]\,=\,$-$0.14, \cite{antipova2003}; HD 116713 and HD 202109 with
respectively [Fe/H]\,=\,$-$0.12 and -0.04, \cite{smiljanic2007}; HD 83548,
[Fe/H]\,=\,+0.03, \cite{pereira2011} ; and NGC 2420\,-\,D, [Fe/H]\,=\,$-$0.55,
\cite{smith1suntzeff987}).  Metal-rich stars have a mean distance from the
Galactic plane of 0.2 kpc and and their Galactic orbits appear to be located
inside the solar Galactic orbit \citep{grenon1999}.  Metal-rich stars in the
solar neighborhood either could be diffused into the disk from the bulge
population \citep{barbuygrenon1990} or they may be the representatives of the
last stages of the chemical evolution of the disk \citep{matteucci2003}.

\par We  here analyze the high-resolution spectra of a sample of 
metal-rich barium stars  to obtain [X/Fe] versus [Fe/H] trends and also
to investigate the heavy-element abundance pattern in metal-rich
environments. As we shall see, all our stars  have
spectral types G and K, and are accordingly   free from the strong molecular opacity
from ZrO, CN and C$_{2}$ absorption features, which complicates a quantitative
analysis and the measurement of some atomic lines. These molecules are usually
observed in MS, S, and disk carbon stars at near solar metallicity where
enhancements from s-process nucleosynthesis have already been reported
\citep{busso2001}.

\section{Observations}

\par The high-resolution spectra of our stars were obtained with the FEROS (Fiberfed Extended Range Optical
Spectrograph) Echelle spectrograph \citep{kauffer1999} at the 1.52\,m and
2.2\,m ESO telescopes at La Silla (Chile) in several runs between 2000 and
2010.  The FEROS spectral resolving power is $R=48\,000$, corresponding to 2.2
pixels of $15\,\mu$m, and the wavelength coverage goes from 3\,800\,{\AA} to
9\,200\,{\AA}.  The nominal $S/N$ ratio was evaluated by measuring the rms
flux fluctuation in selected continuum windows, and the typical values were
$S/N = 100-200$.  The spectra were reduced with the MIDAS pipeline reduction
package consisting of the following standard steps: CCD bias correction,
flat-fielding, spectrum extraction, wavelength calibration, correction of
barycentric velocity, and spectrum rectification.  Table 1 shows the log of
observations and some previous information (V-magnitude, spectral types and
the literature sources from where the stars were selected) of the studied
stars.

\begin{table*} %T1
\caption{Log of the observations and basic information of the stars. }
\begin{tabular}{cccccc}\hline
Star       & Date       & Exposure time & V (mag)  & SpT$^{a}$ & Source \\
           &            &  (sec)       &          &           &        \\\hline
CD-25\,6606& 22/02/2008 &  1\,200       &  9.3     &  ---      &   II$^{b}$ \\
HD 46040   & 14/10/2008 &     420       &  8.0     &   K0p     &    I$^{c}$ \\
HD 49841   & 27/01/2001 &  3\,600       &  9.0     &   G5      &    I   \\
HD 82765   & 17/02/2008 &     600       &  8.5     &   G8\,III &   II   \\
HD 84734   & 17/02/2008 &     600       &  9.0     &   ---     &   II   \\
HD 85205   & 25/03/2000 &     720       &  8.4     &   G6\,II  &   II   \\
HD 100012  & 13/02/2008 &     420       &  6.6     & K0/K1\,III&   II   \\
HD 101079  & 03/04/2007 &     600       &  8.2     &   K2      &  B81$^{d}$ \\
HD 130386  & 28/07/2009 &     420       &  7.8     &   K2      &   II   \\
HD 139660  & 03/04/2007 &     900       &  8.4     &   K0      &  G97$^{e}$ \\
HD 198590  & 18/08/2008 &     360       &  7.0     &   G8\,II  &   II   \\
HD 212209  & 18/08/2008 &     600       &  8.9     &   ---     &   II   \\\hline
\end{tabular}

a: \cite{gomez1997}\par
b: Table II of \cite{macConnell1972}\par
c: Table I of \cite{macConnell1972}\par
d: \cite{bidelman1981}\par
e: \cite{gomez1997}\par
\end{table*}

\section{Analysis and results}

\subsection{Line selection, measurements, and oscillator strengths}

\par Several atomic absorption lines used in this study are the
same as were used in previous studies dedicated to the analysis of photospheric
abundances of the barium stars 
\citep{pereira2005,pereiradrake2009,pereiradrake2011}.
  The chosen lines are sufficiently unblended to yield reliable
abundances. Table~2 shows the Fe\,{\sc i} and Fe\,{\sc ii} lines employed in
the analysis and also the lower excitation potential, $\chi$(eV), of the
transitions, the $\log gf$ values, and the measured equivalent widths.  The
latter were obtained by fitting Gaussian profiles to the observed ones.  The
$\log gf$ values for the Fe\,{\sc i} and Fe\,{\sc ii} lines were taken from
\cite{lambert1996} and \cite{castro1997}.  \addtocounter{table}{1}

\par  Absorption lines whose equivalent widths 
are not stronger than 150\,m\AA\,  should lie in the linear part of the
curve of growth \citep{hill1995}. We can evaluate if this limit is also
valid for our measurements of equivalent widths by analyzing a plot of line
depth vs. equivalent width. In this kind of plot we aim to examine whether
(and where) there would be a significant change in the behavior of the
absorption lines when the weaker and stronger lines are studied together.
Figure 1 shows these measurements of several Fe\,{\sc i} lines in the
spectrum of CD-25\,6606.  We fitted the data using two linear functions, one
between 0.0 m\AA\, and $W_{\mathrm{turn}}$, and another between
$W_{\mathrm{turn}}$ and 300 m\AA. By varying the value of
$W_{\mathrm{turn}}$ between 100 and 200 m\AA\, we verified that the best
fits for the two linear functions simultaneously hold for $W_{\mathrm{turn}}
\sim 160-170$ m\AA. At this point, the slope shows a reduction of about 50\%
(from $\sim 0.0031$ to $\sim 0.0017$). Figure 1 also shows that up to $\sim$
160\,m\AA\, the $\chi$-square of fit is lower (0.0008) than the
$\chi$-square between 160 and 300\,m\AA\,, which is 0.002.  The change of
slope and the large dispersion upward of 160\,m\AA\, are caused by saturation
effects that become important for the stronger lines so that the Gaussian
profiles would not be the best representation for the line profiles.
Therefore, lines with equivalent widths stronger than 160 m\AA\, were not
used in our abundance analysis for the stars analyzed in this work. For the 
same reason we did not use any of the barium lines in our sample stars.

\begin{figure}
\centering
\includegraphics[width=9.5cm]{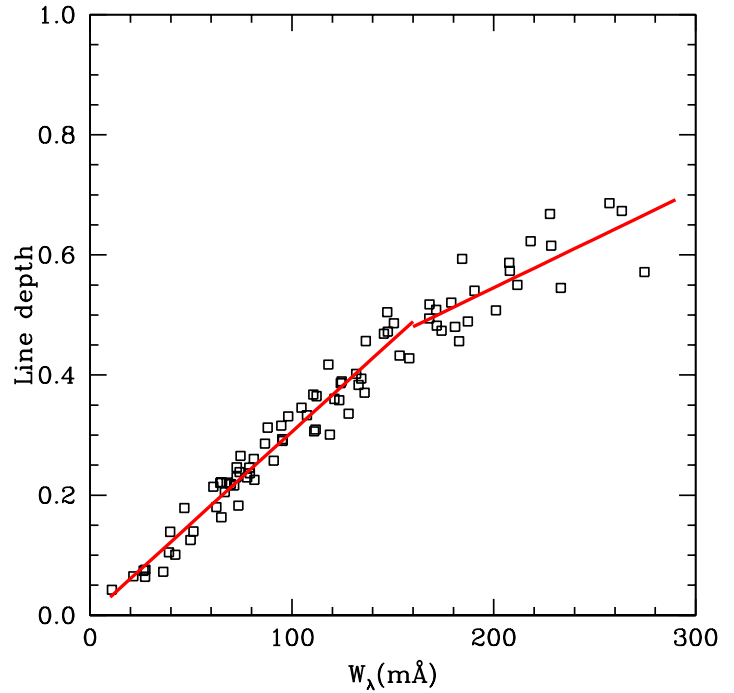}
\caption{Line depth vs. equivalent width of Fe\,{\sc i} lines of CD-25\,6606.
This plot shows that there is a change in the slope around 160\,m\AA\, which 
was used to set this limit for the equivalent widths in our
abundance analysis.}
\end{figure}

\subsection{Determination of the atmospheric parameters}

\par The determination of stellar atmospheric parameters effective
temperature (T$_{\rm eff}$), surface gravity (log $g$), microturbulence
($\xi$), and [Fe/H] (throughout, we use the notation
[X/H]=log(N(X)/N(H))$_{\star}$\,$-$\,log(N(X)/N(H)$_{\odot}$)) are
prerequisites for the determination of photospheric abundances. The
atmospheric parameters were determined by assuming  local thermodynamic
  equilibrium (hereafter LTE) model atmospheres of Kurucz (1993) using the
spectral analysis code MOOG \citep{sneden1973}.

\par The solution of the excitation equilibrium used to derive the
temperature (T$_{\rm eff}$) was defined by the zero slope of the  trend
  between the iron abundances derived from FeI lines and the excitation
  potential of the measured lines.  The microturbulent velocity ($\xi$) was
determined by constraining the abundance determined from individual Fe\,{\sc
  i} lines to show no dependence on $W_{\lambda}/{\lambda}$.  The solution
thus found is unique, depending only on the set of Fe I,II lines and the
atmospheric model employed, and yields as a by-product the metallicity of the
star [Fe/H].  The final adopted atmospheric parameters are given  in
  Table~3.  The value of log $g$ seen in Table 3 was determined by means
  of the ionization balance using the assumption of LTE.  We found typical
uncertainties of $\sigma$(T$_{\rm eff}$) = 150\,K, $\sigma$(log g) = 0.2\,dex,
and $\sigma$($\xi$) = 0.2 km\,s $^{-1}$.

\par Previous determinations of atmospheric parameters of two stars of
our sample, HD 100012 and HD 130386, were made by \cite{luckbond1991} and
\cite{antipova2004}, respectively. \cite{luckbond1991} found for HD 100012,
T$_{\rm eff}$\,=\,5000\,K, log $g$\,=\,2.3 dex, V$_{t}$\,=\,2.2 km\,s$^{-1}$
and [Fe/H]\,=\,+0.33, while for HD 130386, \cite{antipova2004} found T$_{\rm
  eff}$\,=\,4\,720\,K, log $g$\,=\,2.4 dex, V$_{t}$\,=\,1.4km\,s$^{-1}$ and
[Fe/H]\,=\,0.01 dex.

\begin{table*} 
\caption{Atmospheric parameters, radial velocities, and galactic latitude.}
\centering
\begin{tabular}{lccccccc}
\hline\hline
           & $T_{\rm eff}$ (K) & $\log g$  & $[$FeI/H$]$  & $[$FeII/H$]$ 
& $\xi$ (km\,s$^{-1}$) & $V_{r}$ (km\,s$^{-1}$) & latitude gal. ($b^\circ$)\\

CD-25\,6606& $5300$ & $2.7$  & $+0.12\pm0.14$\,(55) & $+0.10\pm0.09$\,(10) &  
$1.5$      & $+0.99\pm0.91$  & $+11.04$ \\

HD 46040   & $4800$ & $2.4$  & $+0.11\pm0.13$\,(27) & $+0.11\pm0.14$\,(9) &  
$1.4$      & $+23.72\pm0.33$ & $-29.59$ \\

HD 49841   & $5200$ & $3.2$  & $+0.21\pm0.13$\,(55) & $+0.20\pm0.19$\,(12) &  
$1.3$      & $+8.17\pm0.50$  & $+02.40$ \\

HD 82765   & $5100$ & $2.6$  & $+0.19\pm0.10$\,(43) & $+0.20\pm0.11$\,(11) &
$1.4$      & $+10.53\pm0.49$ & $-10.65$ \\ 

HD 84734   & $5200$ & $2.9$  & $+0.20\pm0.12$\,(40) & $+0.21\pm0.11$\,(11) &
$1.5$      & $-27.38\pm2.81$ & $-11.68$ \\ 

HD 85205   & $5300$ & $2.8$  & $+0.23\pm0.16$\,(41) & $+0.24\pm0.09$\,(12) &  
$1.9$      & $+25.31\pm0.58$ & $+13.70$ \\

HD 100012  & $5000$ & $2.7$  & $+0.18\pm0.12$\,(45) & $+0.19\pm0.10$\,(11) &  
$1.5$      & $+8.95\pm0.65$  & $+33.59$ \\

HD 101079  & $5000$ & $2.7$  & $+0.10\pm0.12$\,(52) & $+0.09\pm0.08$\,(11) &  
$1.3$      & $-2.77\pm0.55$  & $+56.64$ \\

HD 130386  & $4900$ & $2.7$  & $+0.16\pm0.13$\,(39) & $+0.14\pm0.11$\,(10) &  
$1.3$      & $+16.46\pm0.45$ & $+46.88$ \\

HD 139660  & $5000$ & $2.8$  & $+0.26\pm0.14$\,(44) & $+0.26\pm0.11$\,(12) &  
$1.4$      & $-13.22\pm0.54$ & $+35.11$ \\

HD 198590  & $5100$ & $2.6$  & $+0.18\pm0.14$\,(48) & $+0.17\pm0.13$\,(11) &  
$1.5$      & $-9.62\pm0.42$  & $-39.51$ \\

HD 212209  & $4700$ & $2.4$  & $+0.30\pm0.13$\,(38) & $+0.30\pm0.13$\,(10) &  
$1.2$      & $-27.23\pm0.52$ & $-48.29$ \\\hline
\end{tabular}
\end{table*}

\subsection{Abundance analysis} % 3.3

\par The abundances of chemical elements were determined with
the  LTE model atmosphere techniques.  In brief, equivalent widths are
calculated by integration through a model atmosphere and are compared with the
observed equivalent widths. The calculations are repeated, changing the
abundance of the element in question, until a match is achieved. The current
version of the line-synthesis code {\sc moog} \citep{sneden1973} was used to
carry out the calculations.  Table~4 shows the atomic lines used to derive the
abundances of the elements. Atomic parameters for several transitions of Ti,
Cr, and Ni were retrieved from the library of the National Institute of Science
and Technology Atomic Spectra Database \citep{martin2002}. Tables~5 and 6
provide the derived abundances in the notation [X/Fe].  The last two columns
of Table 6 give the mean abundance of the s-process elements and their
standard deviations, i.e., the scatter around the mean.
\addtocounter{table}{1}

\begin{table*} 
\caption{Abundance ratios [X/Fe] for the elements from Na to Ni.}
%\centering
%\renewcommand{\footnoterule}{}
\begin{tabular}{lcccccccc}
\hline\hline
star    & [Na/Fe]& [Mg/Fe]& [Al/Fe]& [Si/Fe]& [Ca/Fe]& [Ti/Fe]& [Cr/Fe]& [Ni/Fe]
\\\hline
CD-25\,6606 & +0.19 & +0.05 & -0.02 &  +0.11 &  +0.08 &  -0.04 &  +0.03 &  0.00\\  
HD 46040    & +0.16 & +0.04 & +0.02 &  -0.09 &  +0.11 &  +0.03 &  +0.06 & -0.04\\  
HD 49841    & +0.04 & -0.08 & -0.11 &  -0.03 &  +0.07 &  -0.01 &   0.00 & +0.02\\  
HD 82765    & +0.15 & +0.18 & +0.06 &  +0.15 &  +0.14 &   0.00 &  +0.06 & +0.08\\  
HD 84734    & +0.12 & +0.19 & -0.08 &  +0.21 &  +0.06 &  +0.01 &  +0.05 & +0.03\\  
HD 85205    & +0.23 & +0.11 & +0.11 &  +0.09 &  +0.08 &  -0.05 &  -0.01 & +0.08\\  
HD 100012   & +0.29 & +0.19 &  0.00 &  +0.15 &  +0.05 &  +0.04 &   0.00 & +0.06\\  
HD 101079   & +0.20 & +0.14 & +0.23 &  +0.19 &  +0.10 &  -0.06 &  +0.02 & +0.03\\  
HD 130386   & +0.36 & +0.19 & +0.21 &  +0.19 &  +0.15 &  +0.09 &  +0.03 & +0.09\\  
HD 139660   & +0.23 & +0.06 & +0.07 &  +0.19 &  +0.07 &  -0.05 &  +0.03 & +0.05\\  
HD 198590   & +0.21 & +0.20 & +0.16 &  +0.22 &  +0.10 &  -0.09 &  -0.02 & +0.03\\  
HD 212209   & +0.15 & +0.12 & +0.18 &  +0.20 &  +0.06 &  -0.03 &   0.00 & +0.07 \\
\hline
\end{tabular}
\end{table*}

\begin{table*} 
\caption{Abundance ratios [X/Fe] for the heavy elements.}
\begin{tabular}{lccccccc}
\hline\hline
star       &  [Y/Fe] & [Zr/Fe]& [La/Fe] & [Ce/Fe]& [Nd/Fe]& [s/Fe] & $\sigma$([s/Fe])\\
\hline 
CD-25\,6606 &   +0.62 &  +0.48 &  +0.73  &  +0.52 &  +0.85 & +0.64  &  0.15 \\
HD 46040    &   +0.93 &  +1.04 &  +1.83  &  +1.07 &  +0.94 & +1.16  &  0.38 \\
HD 49841    &   +0.85 &  +0.65 &  +0.81  &  +0.56 &  +0.92 & +0.76  &  0.15 \\
HD 82765    &   +0.52 &  +0.37 &  +0.35  &  +0.12 &  +0.20 & +0.31  &  0.14 \\ 
HD 84734    &   +0.64 &  +0.60 &  +0.74  &  +0.56 &  +0.80 & +0.67  &  0.10 \\
HD 85205    &   +0.65 &  +0.65 &  +0.58  &  +0.44 &  +0.80 & +0.62  &  0.14 \\ 
HD 100012   &   -0.03 &  +0.13 &  +0.29  &  +0.25 &  +0.45 & +0.22  &  0.18 \\
HD 101079   &   +0.63 &  +0.44 &  +0.43  &  +0.28 &  +0.51 & +0.46  &  0.13 \\
HD 130386   &   +0.65 &  +0.59 &  +0.47  &  +0.47 &  +0.29 & +0.49  &  0.14 \\
HD 139660   &   +0.77 &  +0.51 &  +0.56  &  +0.35 &  +0.50 & +0.54  &  0.15 \\
HD 198590   &   +0.69 &  +0.51 &  +0.38  &  +0.33 &  +0.33 & +0.45  &  0.15 \\
HD 212209   &   +0.53 &  +0.24 &  +0.24  &  +0.06 &  +0.18 & +0.25  &  0.17 \\\hline  
\end{tabular}
\end{table*}

\subsection{Abundance uncertainties}    %3.4

\par The uncertainties of the derived abundances for the program stars
are dominated by two main sources: the errors in the stellar parameters and
errors in the equivalent width measurements.

\par The abundance uncertainties owing to the errors in the stellar atmospheric 
parameters $T_{\rm eff}$, $\log g$, and $\xi$ were estimated by changing these
parameters by their standard errors and then computing the changes incurred in
the element abundances. This technique was applied in the abundances
determined from equivalent line widths.  The results of these calculations for
CD-25$^\circ$6606 are displayed in columns 2 to 5 of Table 7. The abundance
variations for the other stars show similar values.

\par The abundance uncertainties owing to the errors in the equivalent widths 
measurements were computed from an expression provided by \cite{cayrel1988}.
The errors in the equivalent widths are set, essentially, by the
signal-to-noise ratio and the resolution of the spectra. In our case, having
$R\approx 50\,000$ and typical S/N of 150, the expected uncertainties in the
equivalent widths are about 2--3 m{\AA}. For all measured equivalent
widths, these uncertainties lead to the errors in the abundances less than
those from   the sum of the uncertainties owing to the stellar parameters.

\par Under the assumption that the errors are independent, they can be
combined quadratically so that the total uncertainty is
\[ \sigma = \sqrt{\sum_{i=1}^{N} \sigma^{2}_{i}.} \]
These final uncertainties are given in the sixth column of Table 7.  
The last column gives the observed
abundance dispersion among the lines for those elements with more than three
available lines. Table 7 also shows that neutral elements are fairly sensitive
to the temperature variations, while single ionized elements are sensitive 
to the log $g$ variations. The uncertainties on microturbulence also
contribute to the compounded errors.

\begin{table*}  %Tab 5
\begin{center}
\caption{Abundance uncertainties for CD-25\,6606. The second 
column gives the variation of the abundance  caused by the variation in
$T_{\rm eff}$. The other columns refer to the variations caused by
 $\log g$, $\xi$, and $W_\lambda$, respectively. The sixth column gives
the compounded r.m.s. uncertainty of the second to fifth columns. The
last column gives the observed abundance dispersion for those elements
whose abundances were derived using more than three lines.}
\label{table:5}
\centering
\begin{tabular}{lcccccc}\hline\hline
Species & $\Delta T_{\rm eff}$ & $\Delta\log g$ & $\Delta\xi$ & $\Delta W_{\lambda}$
& $\left( \sum \sigma^2 \right)^{1/2}$ & $\sigma$$_{\rm obs}$ \\
$_{\rule{0pt}{8pt}}$ & $+150$~K & $+0.3$ & +0.3 & +3 m\AA & & \\
\hline                                  
Fe\,{\sc i}    & +0.12  &  0.00 & -0.13 & +0.07 & 0.19 & 0.14 \\
Fe\,{\sc ii}   & -0.09  & +0.11 & -0.14 & +0.07 & 0.21 & 0.09 \\
Na\,{\sc i}    & +0.10  & -0.01 & -0.06 & +0.06 & 0.13 & 0.15 \\
Mg\,{\sc i}    & +0.05  & -0.03 & -0.06 & +0.04 & 0.09 & 0.18 \\
Al\,{\sc i}    & +0.07  & -0.02 & -0.05 & +0.05 & 0.10 & 0.19 \\
Si\,{\sc i}    & +0.01  & +0.02 & -0.05 & +0.06 & 0.08 & 0.06 \\
Ca\,{\sc i}    & +0.13  & -0.01 & -0.14 & +0.07 & 0.20 & 0.10 \\
Ti\,{\sc i}    & +0.28  & +0.01 & -0.07 & +0.07 & 0.29 & 0.09 \\
Cr\,{\sc i}    & +0.15  & +0.02 & -0.09 & +0.08 & 0.19 & 0.18 \\
Ni\,{\sc i}    & +0.11  & +0.03 & -0.10 & +0.07 & 0.17 & 0.12 \\
Y\,{\sc ii}    & +0.01  & +0.12 & -0.16 & +0.08 & 0.22 & 0.06 \\
Zr\,{\sc i}    & +0.21  & +0.02 & -0.01 & +0.08 & 0.23 & 0.04 \\
La\,{\sc ii}   & +0.02  & +0.13 & -0.06 & +0.07 & 0.16 & 0.19 \\
Ce\,{\sc ii}   & +0.02  & +0.13 & -0.09 & +0.08 & 0.18 & 0.11 \\
Nd\,{\sc ii}   & +0.04  & +0.15 & -0.15 & +0.10 & 0.24 & 0.14 \\
\hline
\end{tabular}
\end{center}
\end{table*}

\section{Discussion}    %4

\subsection{The position of the stars in the log\,$g$\,-\,log\,T$_{\rm eff}$
diagram.}

\par Figure~2 shows the position of the studied stars in the $\log T_{\rm
  eff}-\log g$ diagram, where model tracks were computed by Schaerer et al.
(1993) for the stars of 2.0, 2.5, and 3.0 solar masses with metallicity of
Z\,=\,0.04.  Putting the studied stars in this diagram, we estimated the masses
as given in Table 8.  Table 8 also provides the distance that resulted from
the assumed mass from the $\log T_{\rm eff}-\log g$ diagram and from the
temperature and surface gravity given in Table 3. In these calculations we
assumed the bolometric corrections of \cite{alonso1999} and M$_{\rm bol\odot}$
= +4.74 \citep{bessell1998}.  The last column of Table 8 provides the distance
of the stars whose parallax were measured by Hipparcos.  The distances were
derived based on the ionization equilibrium and the derived masses from the
positions in the $\log g-\log T_{\rm eff}$ diagram, and astrometric values agree well
within the uncertainties with the  possible exception of HD 13966.

\par The stars CD-25\,6606, HD 49841, HD 84734, and HD 85205, 
with log\,T$_{\rm eff}$\,=\,3.72 and log\,g\,=\,2.7, 3.2, 2.9, and 2.8
respectively, occupy a region that is at the end of subgiant phase or near
the base of red giant branch in the $\log g-\log T_{\rm eff}$ diagram.  
This is important because in all high-resolution spectroscopy studies
dedicated to the analyses of the spectra of barium dwarfs and CH subgiant
stars, no stars of near solar metallicity or even higher than the solar
metallicity in this evolutionary phase were  found
\citep{smith1993,pereira2005,allenbarbuy2006}.  Therefore, these four stars
might be the first identified metal-rich CH subgiants. Barium dwarfs (log $g$
$\geq$ 4.0) with metallicities close to the solar one are as yet undiscovered
\citep{pereiradrake2011}.

\begin{figure}
\centering
\includegraphics[width=9.5cm]{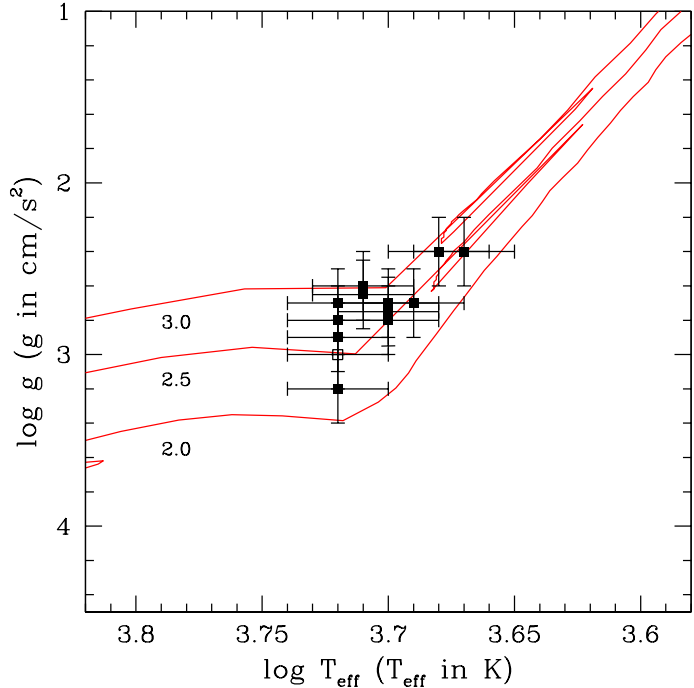}
\caption{Position of the stars (filled squares) in the 
$\log g-\log T_{\rm eff}$ diagram. Evolutionary tracks are from 
\cite{schaerer1993}. The numbers correspond to stellar masses in units of 
solar mass. Red lines represent evolutionary tracks for a star with Z\,=\,0.04.
The stars  CD-25\,6606, HD 49841, HD 84734, and HD 85205 are subgiant stars
about to begin  their ascent of the red giant branch, like S\,190, 
a D$^{'}$-type symbiotic star ({\sl open square}) that is also barium 
enriched \citep{smith2001}.}
\end{figure}

\begin{table} 
\caption{Masses and distances of the  stars.}    
\begin{tabular}{lccc}\hline                               
star       &  M/M$_{\odot}$ & r (pc) & r (pc)(Hipparcos) \\\hline
CD-25\,6606& 3.0 & 820$\pm$200 &  --- \\
HD 46040   & 2.5 & 440$\pm$100 & 320$\pm$70 \\
HD 49841   & 2.0 & 310$\pm$70  &  ---  \\
HD 82765   & 2.5 & 520$\pm$120 & 420$\pm$140 \\
HD 84734   & 2.5 & 490$\pm$115 &  ---  \\
HD 85205   & 2.5 & 440$\pm$100 &  ---  \\
HD 100012  & 2.5 & 190$\pm$45  & 180$\pm$25 \\ 
HD 101079  & 2.5 & 390$\pm$90  &  ---  \\
HD 130386  & 2.5 & 300$\pm$70  & 290$\pm$110 \\
HD 139660  & 2.5 & 380$\pm$90  & 190$\pm$44  \\
HD 198590  & 2.5 & 260$\pm$60  & 410$\pm$140 \\
HD 212209  & 2.5 & 630$\pm$150 & 390$\pm$180 \\\hline
\end{tabular}
\end{table}

\subsection{Kinematics}

\par As we mentioned in the introduction,  some metal-rich stars could
  be old objects.  Figure 3 shows the position of stars analyzed in this work
in the HR-diagram with the isochrones of \cite{girardi2000}.  The metal-rich
barium stars with ages between 8.5 and 9.0 Gyr are similar to the metal-rich
stars analyzed by \cite{pakhomov2009}, which have a  mean age of
$\approx$\,8.7\,Gyr.  Our results confirm the results of \cite{mennessier1997}
that some barium stars in this range of gravity (and also in absolute 
magnitudes) have ages between 8.5 and 9.5 Gyr.

\begin{figure}
\centering
\includegraphics[width=9.5cm]{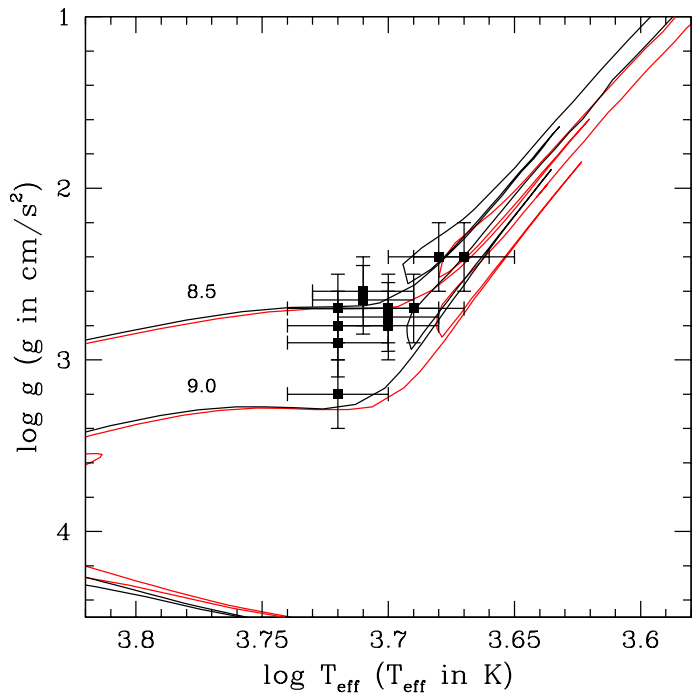}
\caption{Positions of the metal-rich barium stars compared with isochrones of 
\cite{girardi2000}. Red lines represent metallicities of Z\,=\,0.03 and
black lines represent metallicities of Z\,=\,0.02. The numbers correspond to
the age in units of Gyr.}
\end{figure}

\par We also investigated the kinematical properties of our sample stars.
Table 9 shows our results. The radial velocities were determined by measuring
the Doppler shift of the spectral lines and are given in Table 3. Distances
and proper motions were taken from  the Hipparcos catalog  \citep{perryman1997}.
 Space velocities (U$_{\odot}$, V$_{\odot}$, W$_{\odot}$) relative to the
  local standard of rest were computed based on the method of
  \cite{johnsonsoderblom1987} and were calculated assuming the solar motion of
  (U, V, W)\,=\,(11.1, 12.2, 7.3) km\,s$^{-1}$, as derived by
  \cite{schonrich2010}.  The Galactic orbital parameters R$_{\sl min}$ and
R$_{\sl max}$ (minimum and maximum distances from the Galactic center),
Z$_{\sl max}$ (maximum distance from Galactic plane) and $e$ (orbital
eccentricity) were obtained using the  Galactic potential integrator
developed by C. Flynn (http://www.astro.utu.fi/galorb.html). For this we
adopted a solar Galactocentric distance of 8.5 kpc and circular velocity of
220 km\,s$^{-1}$.  Because  barium stars are binary stars, their observed radial
velocities are affected by periodical variations. If we have only one
determination of radial velocity, it is interesting to evaluate the
uncertainties introduced by an unknown orbital motion.  For this purpose we
analyzed observed orbital velocity amplitude of barium binary stars with
determined orbital elements \citep{jorissenmayor1988,udry1998a,udry1998b}.
Figure 4 shows the dependence between observed orbital velocity amplitudes and
the periods of analyzed stars. The mean value of the amplitude for the whole
sample was found to be 6 km\,s$^{-1}$. This value was considered as the
uncertainty in the determined radial velocities of our stars.

\begin{figure}
\centering
\includegraphics[width=9.0cm]{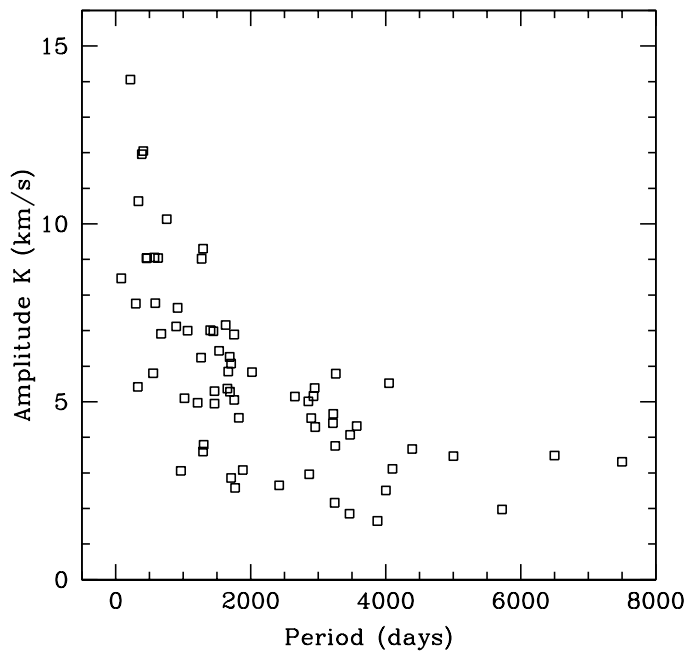}
\caption{Amplitude $versus$ period for a sample of barium stars.}
\end{figure}

\begin{table*} 
\caption{Space velocities and Galactic orbital parameters for some metal-rich 
barium stars.} 
\begin{tabular}{|l|c|c|c|c|c|c|c|}\hline                               
& U$_{LSR}$ & V$_{LSR}$  & W$_{LSR}$ & R$_{\sl min}$ & R$_{\sl max}$
& Z$_{\sl max}$ & $e$\\\hline
star       & km\,s$^{-1}$ & km\,s$^{-1}$ & km\,s$^{-1}$ &  kpc & kpc & kpc &\\\hline
HD 46040   &   -2.3 &  -2.2 & -15.0 & 7.19 & 7.95 & 0.3 & 0.05 \\ 
HD 82765   &    9.5 &  -3.7 &   5.9 & 7.73 & 8.12 & 0.1 & 0.02 \\
HD 100012  &   11.8 &  -5.1 &   4.1 & 7.53 & 8.10 & 0.1 & 0.04 \\
HD 130386  &  -20.8 & -25.0 &  13.7 & 6.46 & 8.35 & 0.3 & 0.13 \\
HD 139660  &    5.5 & -26.0 &   4.7 & 6.47 & 8.20 & 0.1 & 0.12 \\
HD 198590  &  -14.9 &   2.8 &  26.3 & 7.97 & 8.72 & 0.6 & 0.05 \\
HD 212209  &   50.6 &  20.0 &  -7.0 & 7.60 & 10.37 & 0.4 & 0.15 \\\hline
\end{tabular}
\end{table*}

\par Inspecting Table 9 we may conclude that our sample stars come 
from several places of the Galaxy, with mean space velocities of
U$_{LSR}$\,=\,6$\pm$23, V$_{LSR}$\,=\,$-$6$\pm$16 and W$_{LSR}$\,=\,5$\pm$13
km\,s$^{-1}$. Their minimum and maximum distances to the Galactic center are
7.3$\pm$0.6\,kpc and 8.5$\pm$0.8\,kpc, respectively.  Most of the stars have
V$_{LSR}$ negative results, which can be taken as indication that they lag in
the Galactic rotation compared to the solar motion.  Two stars, HD 198590 and
HD 212209, have V$_{LSR}$ opposite velocities compared to the rest of the
stars. These properties can be taken as  evidence that these metal-rich
barium stars as well as other metal-rich and super metal-rich stars form an
 inhomogeneous group. Metal-rich stars with space velocities and/or
R$_{\sl min}$ and R$_{\sl max}$ that differ from each other of the same
studied sample would have a different star-formation history and evolution
\citep{pakhomov2009}.  It also interesting to notice that HD 212209 has the
highest eccentricity among the other stars. Most of the stars have R$_{\sl
  max}$ less than 8.5 kpc, that is, within the solar circle, which seems to 
agree with the results obtained by \cite{raboud1998} and
\cite{grenon1999} for metal-rich stars. Other results may be seen in Table
9. All metal-rich barium stars lie inside in the (U,V)-plane of metal-rich
stars of \cite{raboud1998} and \cite{grenon1999}, thus indicating that they
display similar kinematical properties of non s-process enriched metal-rich
stars. Finally, the star HD 198590 with the highest
Z$_{\sl max}$ value among the analyzed stars, with a metallicity of
$\approx$+0.2, lies in the upper limit (with very few others) of the Z$_{\sl
  max}$ value in the Z$_{\sl max}$-metallicity diagram of Grenon (1999). In
Figure 13 of \cite{soubirangirard2005} this star lies between the thin-disk
stars at ([Fe/H],Z$_{\sl max}$)\,=\,(0.0, 0.1-0.2\,kpc) and the thick-disk
stars at ([Fe/H],Z$_{\sl max}$)$\approx$(0.5,1.0 kpc). The other stars in
Table 9 belong to the thin-disk stars with lower Z$_{\sl max}$ values.

\subsection{Abundances}

\par In this section we compare the abundance ratios
[X/Fe] of the metal-rich barium giants analyzed in this work with  the  results
  of some studies already made for giant stars.  We used for comparison the
data from local disk field giants studied by \cite{luckheiter2007}, disk
red-clump giants studied by \cite{mishenina2006,mishenina2007}, metal-rich
giants studied by \cite{pakhomov2009} and two metal-rich open clusters and
$\mu$ Leo studied by \cite{carretta2007}.  Figures 5 to 8 show the [X/Fe]
ratios with metallicity for the barium giants of our sample and the results
for the non s-process enriched giant stars  mentioned above. We will see that
with respect to the sodium, aluminun, $\alpha$-elements,  chromium and
  nickel, the abundance ratios follow the same trend as  seen in the 
disk-giant stars. For the heavy-elements, the [s/Fe] ratio, where 's' means
the mean abundance of the elements synthesized by s-process, ranges from +0.2
to +1.2, thus indicating different degrees of enhancements.

\subsubsection{Sodium to nickel} 

\par Sodium and aluminum are mainly produced by hydrostatic carbon burning in
massive stars \citep{woosleyweaver1995}. Sodium and aluminun in disk stars has
been observed among others by \cite{edvardsson1993} and
\cite{feltzinggustafsson1998}.  Over the range $-$1.0 $<$ [Fe/H] $<$ 0.0, the
ratios [Na/Fe] and [Al/Fe] are slight enhanced by $\approx$ 0.1\,dex. For
metal-rich stars \cite{edvardsson1993} and \cite{feltzinggustafsson1998} (see
also Shi et al., 2004) observed an upturn in the [Na/Fe] ratio for metal-rich
stars, that is [Na/Fe]\,=\,+0.2 at [Fe/H]\,=\,+0.2. \cite{shi2004} argued that
 to explain this Na overabundance for the metal-rich stars, large
amounts of sodium have been produced by NeNa cycle in the interiors of AGB
stars. Figure 5 shows that both the local giants and the red clump giants
exhibit the same trend as seen in dwarf stars. The barium giants of our sample
follow the same trend. For aluminun the ratio [Al/Fe] is almost constant at
$\approx$0.1 in the metallicity range $-$2.0 to 0.2
\citep{carretta2002,edvardsson1993,feltzinggustafsson1998}.  The local giants
and our data for the barium giants  seem to follow the same trend as seen
for all these previous studies.

\par In the thick- and thin-disk stars the $\alpha$-element abundances 
given by the mean of Mg, Si, Ca, and Ti are overabundant by $\approx$ 0.2 dex
at $-$1.0 $<$ [Fe/H] $<$ $-$0.5 and then  decrease by 0.1-0.0 dex at
$-$0.5 $<$ [Fe/H] $<$ 0.0 \citep{edvardsson1993,reddy2003,reddy2006}.
\cite{feltzinggustafsson1998} observed that [$\alpha$/Fe] ratio has the same
trend as  detected by the other authors at $-$0.5 $<$ [Fe/H] $<$ 0.0.  Up
to [Fe/H]\,=\,+0.3 the [$\alpha$/Fe] ratio remains flat at [Fe/H]\,=\,0.0-0.1.
The analysis of the local disk field giants studied by \cite{luckheiter2007}
and disk red-clump giants studied by \cite{mishenina2006} showed that giants
display the same trend as mentioned above  for the dwarfs. Figure 6 shows that
the [$\alpha$/Fe] ratio {\it versus} [Fe/H] for the barium stars follows the
same trend as the giants.  The ratios [X/Fe] for the $\alpha$-elements, Mg,
Si, Ca, and Ti are discussed below.

\par Magnesium (like oxygen) is produced in massive stars at
$\approx$\,25\,M$_{\odot}$ as predicted by the  nucleosynthesis theory 
\citep{woosleyweaver1995} and iron is mainly produced  by SNeI events. It
would be expected that both ratios, [O/Fe] and [Mg/Fe], would decrease with
increasing metallicity.  Indeed, the [O/Fe] ratio at [Fe/H]\,=\,$-$1.0 is
$\approx$+0.4, while at [Fe/H]\,=\,0.2 is $\approx$$-$0.1
\citep{edvardsson1993}.  In halo stars, the [Mg/Fe] ratio also increases (like
oxygen) toward lower metallicities \citep{norris2001}. In dwarf disk stars,
for metallicities higher than [Fe/H] $>$ $-$0.2 and up to +0.2, the ratio
[Mg/Fe] becomes flat at $\approx$0.1 \citep{feltzinggustafsson1998,chen2000}.
This observed trend of the [Mg/Fe] ratio toward the higher metallicities
would suggest that Mg would not only be produced by massive stars but also by
 type I SNe  events \citep{chen2000}.  \cite{timmes1995} predict subsolar
values for the [Mg/Fe] ratio at higher metallicities for the evolution of the
magnesium abundances in the Galaxy, which is not seen in all the analysis of
the [Mg/Fe] ratio. The analysis of the [Mg/Fe] ratio of local giants by
\cite{luckheiter2007} as well as the barium giants of our sample shows the
same trend as seen  in dwarfs for metallicities higher than [Fe/H] $>$
0.0 (Figure 5). However, the [Mg/Fe] ratio for the red clump giants analyzed by
\cite{mishenina2006} behaves like the [O/Fe] ratio.

\par Silicon can be produced by 10-30\,M$_{\odot}$ stars by hydrostatic oxygen
burning and also during the eventual type II supernovae explosion
\citep{woosleyweaver1986}.  In halo stars at [Fe/H]\,=\,$-$3.0 it is enhanced
by $\approx$+0.5 with larger scatter \citep{norris2001,carretta2002}.  For
dwarfs with metallicities higher than [Fe/H]\,=\,0.0, it seems that silicon
behaves like magnesium and has a [Si/Fe] ratio of $\approx$\,0.05
\citep{edvardsson1993}.  Inspecting the results of
\cite{feltzinggustafsson1998},  one can note that silicon flattens at a mean
ratio of [Si/Fe]\,=\,0.0 for [Fe/H]\,=\, 0.0-0.4.   For the red clump
  giants, the local giants and the barium giants analyzed in this work the
  relative-to-iron silicon abundance flattens at a mean ratio of  of
[Si/Fe]$\approx$0.1 for [Fe/H]\,=\, 0.0-0.3.

\par Calcium shares the same nucleosynthesis site as silicon. In halo stars
 it is  enhanced by 0.2-0.3 dex \citep{norris2001}. In dwarfs and for
metallicities higher than [Fe/H] $>$ $-$0.2, calcium flattens toward  a mean
ratio [Ca/Fe] of $\approx$0.0. The same trend is seen in barium and red clump
giants but for the local giants, the [Ca/Fe] ratio is lower by $-$0.2 dex at
[Fe/H]\,=\,+0.2 \citep{luckheiter2007}.

\par Titanium is usually considered an $\alpha$-element that is produced
by explosive oxygen burning but could also be produced by Type Ia supernovae
\citep{woosleyweaver1995}. Observationally, it behaves like an
$\alpha$-element.  Indeed, the [Ti/Fe] ratio for halo stars  displays
a similar trend as the [Ca/Fe] ratio.  In dwarf-disk stars [Ti/Fe]  slightly
  decreases with increasing metallicity \citep{edvardsson1993}. For higher
metallicities it seems that the ratio [Ti/Fe] flattens around $\approx$0.1. In
the present work, the [Ti/Fe] ratios are usually lower than the [Mg,Si,Ca/Fe]
ratios, displaying a similar trend as local-disk giants \citep{luckheiter2007}.

\par The iron peak elements are formed in large amounts in Type Ia supernovae
and all its members should follow the same trend with iron abundance,
therefore the [X/Fe] ratios should remain constant within the range of
metallicity studied. Indeed, nickel, does remain constant with [Ni/Fe]\,=\,0.0 for
$-$0.8 $<$ [Fe/H] $<$ 0.1. \cite{luckheiter2007} also found a tight
correlation between [Fe/H] and [Ni/H] with a slope of 1.04 also between $-$0.8
$<$ [Fe/H] $<$ 0.1. However, some studies show a possible upturn after
[Fe/H]$>$0.1 for red clump giants \citep{mishenina2006,liu2007} and dwarfs
\citep{chen2000}. This upturn is seen in Luck \& Heiter's data and also in
Figure 4 of \cite{soubirangirard2005}.  For the element chromium, the study
for the local giants of \cite{luckheiter2007} found a tight correlation
between [Fe/H] and [Cr/H] with a slope of 1.06, that is [Cr/Fe]\,=\,0.0 for
$-$0.8 $<$ [Fe/H] $<$ 0.2, despite some discrepant points, as they reported.
 The barium stars of our sample also follow the general trend seen for
the field giants, for the ratios of [Ni/Fe] and [Cr/Fe].

\begin{figure}
\centering
\includegraphics[height=11cm,width=9.6cm]{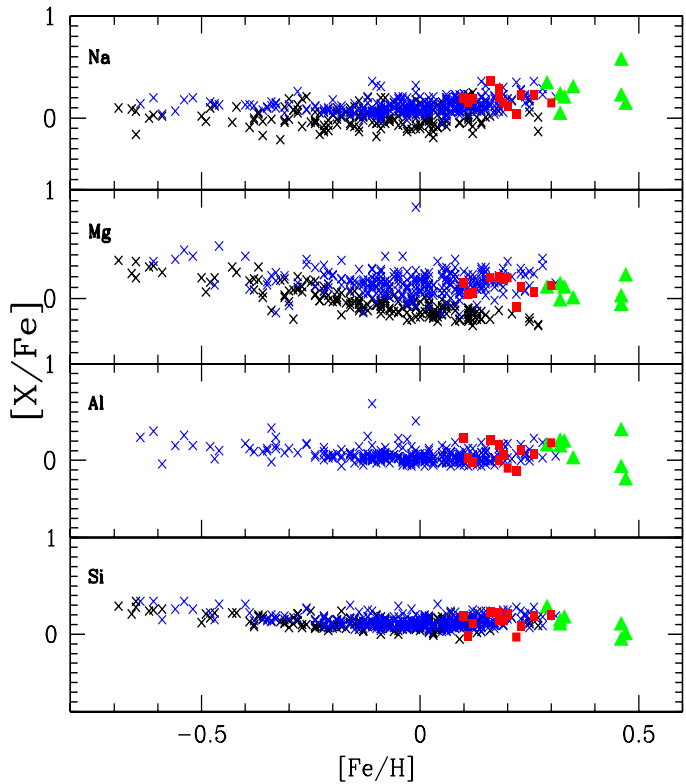}
\caption{Abundance ratios [X/Fe] vs. [Fe/H]. Metal-rich barium stars 
{\sl red squares}; field giants of \cite{luckheiter2007}, {\sl blue crosses};
clump giants of \cite{mishenina2006}, {\sl black crosses};  metal-rich 
giants stars and two metal-rich open clusters and $\mu$ Leo studied
 by \cite{pakhomov2009} and \cite{carretta2007}, respectively, {\sl green triangles}.}
\end{figure}

\begin{figure}
\centering
\includegraphics[width=9.0cm]{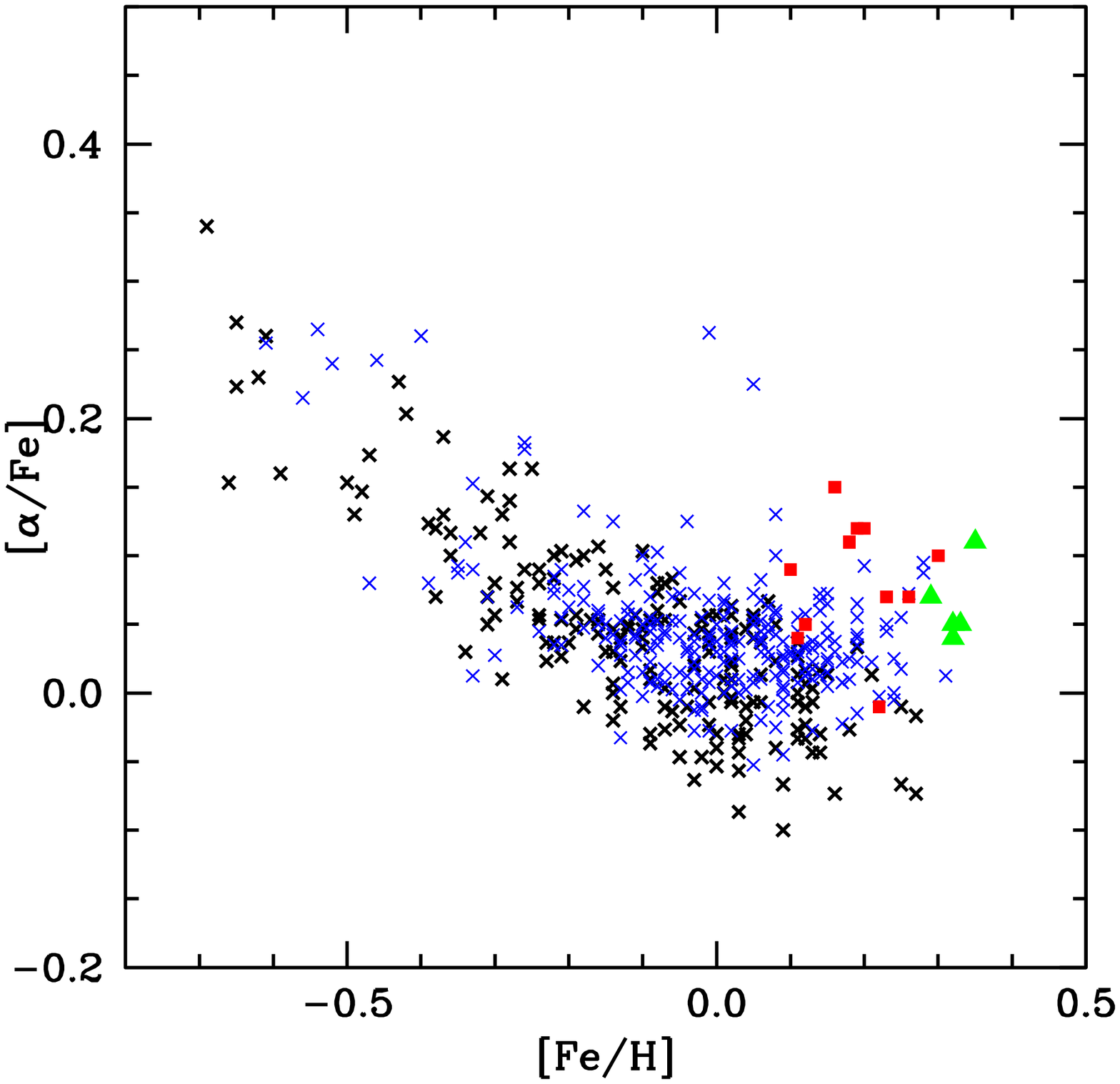}
\caption{Abundance ratios [$\alpha$/Fe] vs. [Fe/H]. Symbols have the same 
meaning as in Fig. 5}
\end{figure}

\begin{figure}
\centering
\includegraphics[height=11cm,width=9.6cm]{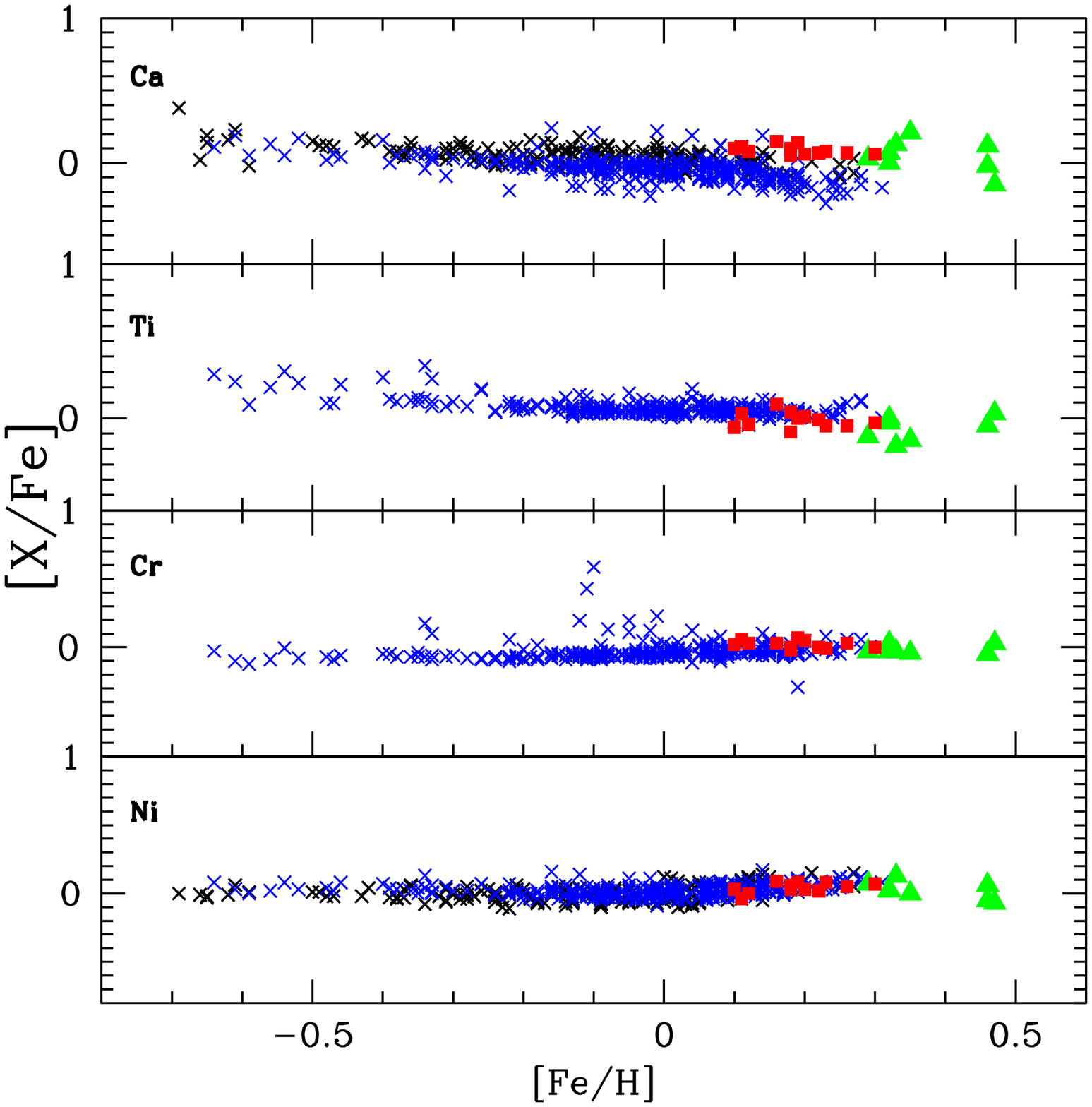}
\caption{Abundance ratios [X/Fe] vs. [Fe/H]. Symbols have the same meaning as
  in Fig. 5}
\end{figure}

\begin{figure}
\centering
\includegraphics[height=11cm,width=9.6cm]{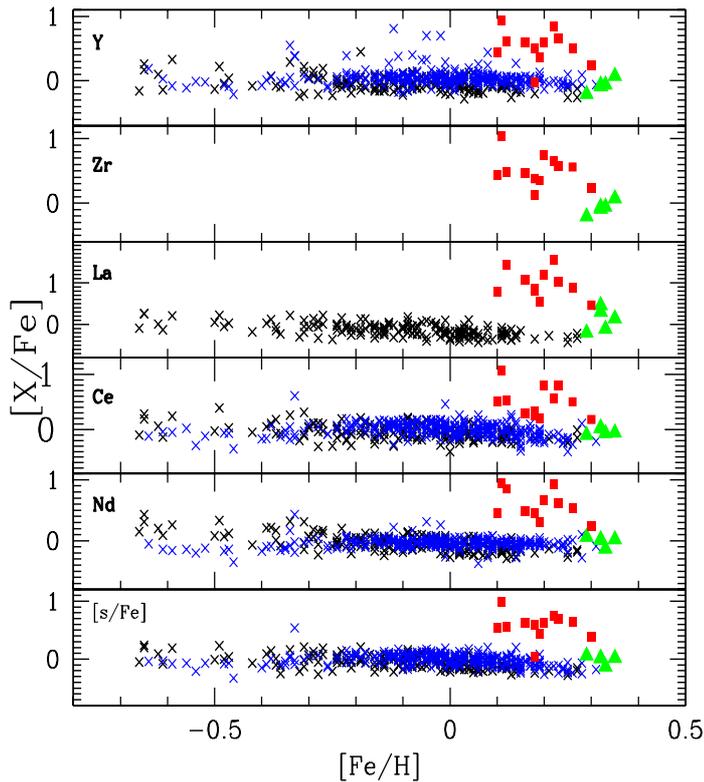}
\caption{Abundance ratios [X/Fe] vs. [Fe/H]. Symbols have the same meaning as
 in Fig. 5. Clump giants are from \cite{mishenina2007}.}
\end{figure}

\subsubsection{s-process elements : Y, Zr, La, Ce and Nd}

\par The main site of $s$-process elements (and also carbon) production are 
  low-mass AGB stars (M$\leq$\,3.0 M$_{\odot}$) during the thermally pulsing
  phase (TP-AGB). Nucleosynthesis of s-process takes place in the C-rich
  intershell convective zone, the zone between the H-burning shell and the
  He-burning shell, of which the source of neutrons for the s-process to occur
  is the $^{13}$C($\alpha$,n)$^{16}$O reaction \citep{lattanziowood2003}.
Figure 8 shows the ratio [s/Fe], where 's' represents the mean of the elements
created by slow neutron capture reactions (s-process): Y, Zr, La, Ce, and Nd.
As we can see from Figure 8, all stars except HD 100012 are enhanced in
the s-process elements when compared to the same ratio of the field stars. HD
100012 has a [s/Fe] ratio of +0.22  and therefore cannot be considered as
  a barium star because some stars at this metallicity  have a similar
[s/Fe] ratio as HD 157919 with [s/Fe]\,=\,+0.2 at [Fe/H]\,=\,+0.14. Among
the local giants analyzed by \cite{luckheiter2007},  only one barium star was spotted,
 HD 104979 \citep{zacs1994}, which can be seen at [Fe/H]\,=\,$-$0.33 with
[s/Fe]\,=\,+0.54.

\par  Figure 8 also shows that our sample stars differs from the 
  field stars only bye  the degree of s-process enrichment observed in their
  atmospheres.  Models of galactic chemical evolution do not predict this
  enrichment at this metallicity \citep{travaglio1999, travaglio2004}.
  Therefore, the atmospheres of these stars were contaminated either through the
  stellar evolution or  an {\sl extrinsic} event that may have happened
  in the past, i.e., mass-transfer hypothesis.  The first hypothesis can be
  ruled out because of their position in the log\,$g$\,-\,log\,T$_{\rm eff}$
  diagram (Figure 2). Finally, Figure 8 also shows that the abundance of zirconium
  is poorly investigated in normal giant stars in this metallicity range.

\par Figure 9 shows the [s/Fe] ratios for  chemically peculiar
binary systems including not only the barium stars (giants and dwarfs) but
also the yellow symbiotic stars, the CH stars, and the  CEMP (carbon
  enhanced metal-poor) stars, those that have already been proved as binary
systems.  The definition of `s' varies from author to author and in some cases
depends on the quality and/or on the wavelength range of the available
spectra.  Among CEMP stars, there are six  that are binaries:
CS~22942-019, CS~22948-027, CS~29497-030, CS~29497-034, CS~22964-161, and
HE~0024-2523, the data of which concerning their carbon and heavy-element (Z
$>$ 56) overabundances and binarity were taken from the results of
\cite{prestonsneden2001,sivarani2004,barbuy2005,lucatello2003,
thompson2008,aoki2002} and \cite{hill2000}.  
According to the mass-transfer hyphotesis for the origin
of the barium stars, one of the binary components ejects its envelope that is enriched
in the s-process elements produced during the AGB phase, and this matter falls
onto the companion, which in turn becomes enriched in these elements, and now is  the barium
star. The different ratios [s/Fe] observed among the barium stars, also seen
in this study, depend on the material received by the companion as well as
whether this material is partially or fully mixed in the atmosphere of the
barium stars.  Theoretical studies for the production of the s-process
elements in intrinsic AGB stars \citep{gallino1998,busso1999} indicate that
there is a trend that the s-process elements are more easily produced at lower
metallicities as a result of the operation of the reaction
$^{13}$C($\alpha$,n)$^{16}$O as a neutron source. When several classes of
chemically-peculiar stars are examined all together (Figure 9), this
phenomenon is observed.

\begin{figure}
\centering
\includegraphics[height=11cm,width=9.6cm]{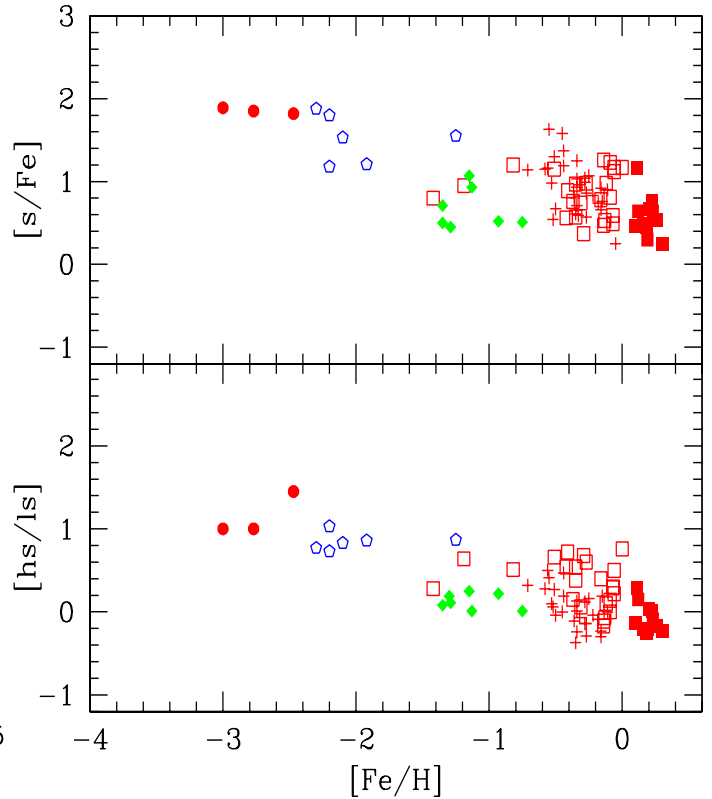}
\caption{Diagram of [s/Fe] {\it versus} [Fe/H] (top)
and [hs/ls] {\it versus} [Fe/H] (bottom)
for several classes of chemically peculiar binary stars.
Metal-rich barium stars ({\it filled red squares});
barium giants previously  analyzed ({\it red open squares});
barium dwarfs ({\it plus red crosses});
CH stars ({\it blue open polygons});
yellow symbiotic ({\it green symbols})
and CEMP-s stars which are {\it members} of binary systems 
({\it red filled circles})}
\end{figure}

\par According to \cite{busso1999} and \cite{travaglio2004}, the 
neutron-capture nucleosynthesis in AGB stars is metallicity-dependent. Among
the elements created by the s-process nucleosynthesis, the first neutron magic
peak elements (such as Y and Zr) are the dominant products of the neutron
captures in AGB models at metallicities around [Fe/H]\,=\,$-$0.4. At lower
metallicities, [Fe/H]$\approx$$-$1.0, the second neutron magic peak elements
(Ba-La-Ce-Nd) are the dominant. The origin of these two peaks is the result of
the neutron magic numbers 50 and 82 nuclei, which have smaller absorption 
  cross-sections than their neighbor nuclei and therefore  turn out  to be more
stable than others.  Owing to this, it is possible to monitor the s-process
efficiency through the relative abundances of the second-peak elements to the
first-peak elements. In this way the s-elements at the second-peak are usually
referred   to as the ``heavy'' s-process elements, in our case, La, Ce, and Nd, while
s-elements at the first-peak are usually referred  to as the ``light'' elements, in our
case, Y and Zr. Thus, the average logarithmic ratio [hs/ls] has been widely
used to measure the efficiency of neutron captures.  Table 10 provides the
ratios [ls/Fe], [hs/Fe] as well as the ratio [hs/ls] for the studied stars.
Inspecting Table 10, we see that our derived values for the  indices
[hs/ls], [hs/Fe], [ls/Fe] for all  metal-rich barium stars, except for HD
46040, are in the range, $-$0.3 to 0.0, 0.2 to 0.7 and 0.4 to
0.8.  These values  seem to agree well  when compared with
model predictions given by \cite{busso2001} as seen in their Figures 7, 8, and
9 for the ratios [hs/ls], [hs/Fe], [ls/Fe] vs metallicity for the
standard case (ST/1.5). The ST case corresponds to
4$\times$10$^{-6}$\,M$_{\odot}$ of $^{13}$C for AGB stars between 1.5 and 3.0
M$_{\odot}$ for a metallicity of [Fe/H]\,=\,$-$0.3, which accounts for the main
component of the s-process in solar system \citep{busso2001}.  HD 46040
has higher values for the  indices [hs/ls] and [hs/Fe] for a star at this
metallicity and its [ls/Fe] index marginally falls in the limit of the model
predictions.

\begin{table} 
\caption{Indices of  the s-process.}    
\begin{tabular}{lccc}\hline                               
star          & [ls/Fe] & [hs/Fe] &  [hs/ls]\\\hline
CD-25\,6606   &  0.55   &  0.70   &   0.15\\
HD\,46040     &  0.99   &  1.28   &   0.29\\
HD\,49841     &  0.75   &  0.76   &   0.01\\
HD\,82765     &  0.43   &  0.22   &  -0.21\\
HD\,84734     &  0.62   &  0.66   &   0.04\\
HD\,85205     &  0.70   &  0.61   &  -0.09\\
HD\,101079    &  0.54   &  0.41   &  -0.13\\
HD\,130386    &  0.62   &  0.41   &  -0.21\\
HD\,139660    &  0.64   &  0.47   &  -0.17\\
HD\,198590    &  0.60   &  0.35   &  -0.25\\
HD\,212209    &  0.39   &  0.16   &  -0.23\\\hline
\end{tabular}
\end{table}

\par In Figure 9 we show the [hs/ls] index  as a function of the 
metallicity for the same classes of chemically peculiar binary systems.  As for
the [s/Fe] ratio vs metallicity, the [hs/ls] index also increases
toward lower metallicities, with a large spread. In both figures we can see
that the position of metal-rich barium stars are useful targets for constraining
models for the nucleosynthesis of the s-process at higher metallicities.

\section{Conclusions}

\par  Our abundance analysis has been performed by employing 
high-resolution optical spectra of a sample of metal-rich barium stars
with the aim to obtain their abundance pattern and kinematical properties,
and it can be summarized as follows:

\begin{enumerate}
  
\item The stars have enhancement factors, [s/Fe], from 0.25 up to 1.16.  These
  overabundances are expected for stars at this metallicity according to the models
  of \cite{busso2001}, which are based on the observed  indices [ls/Fe],
  [hs/Fe] and [hs/ls].  The abundance pattern of the iron peak elements, of
  the elements of the $\alpha$ group, as well as that of sodium and aluminum
  follows the disk abundance.   The [alpha/Fe] ratios in our stars do not
    show any trace of overabundances, while bulge stars display increased
    ratios \citep{lecureur2007}. Therefore, these metal-rich barium stars do
  not belong to the bulge population. It is probable that like the most of the
  non s-process enriched metal-rich stars, the barium stars analyzed in this
  work belong to the thin-disk stars, as can be deduced form their low Z$_{\sl max}$
  values. Also interesting to note is that all stars analyzed in this work
  have low values for the eccentricities, which  indicates almost circular 
    orbits in agreement with this characteristic of the other non s-process
  enriched metal-rich stars previously analyzed
  \citep{pakhomov2009,grattonsneden1990} and which agrees with the results of
  \cite{raboud1998}.

\item Considering evolutionary aspects, metal-rich barium stars put another
  constraint on the evolution of barium dwarfs to barium giants. Because all 
  CH subgiants and barium dwarfs analyzed  so far have metallicities less
    than the  metallicity of the Sun, except those very few subgiants analyzed in
  this paper, metal-rich barium dwarfs/subgiants are yet to be found. In
  this respect the same question can be addressed to the yellow symbiotics.
  All yellow symbiotics, except for the D$^{'}$-type, which probably are not
  symbiotic stars \citep{corradischwartz1997,pereira2005}, are halo objects
  and their connection with metal-poor barium stars have been investigated in
  the recent years \citep{smith1996,smith1997,jorissen2005,pereiraroig2009}.
  Consequently, disk yellow symbiotics at near solar metallicity have not yet been
  identified.
  
\item Metal-rich barium stars are useful objects for investigating the
  heavy-element abundance pattern  at metalicities higher than the
    metallicity of the Sun because they are free of many molecular
    features that are usually seen in the spectra of MS, S, and carbon stars. The
  barium star phenomenon once it extend toward higher metallicities, shares
  several properties with the non s-process enriched metal-rich and super
  metal-rich giant stars previously investigated.

\end{enumerate}

\begin{acknowledgements}  This research has made use of the SIMBAD 
database, operated at CDS, Strasbourg, France. We thank our referee for 
detailed comments that have improved the presentation of the present work.

\end{acknowledgements}

\longtab{2}{
\begin{longtable}{ccccccccccc}
\caption{Observed Fe\,{\sc i} and Fe\,{\sc ii} lines.}\\
\hline\hline
   & & & & \multicolumn{7}{c}{Equivalent Widths (m\AA)}\\
\cline{5 - 11}
   & & & & CD & HD & HD & HD & HD & HD & HD\\
Element &  $\lambda$\,(\AA) &  $\chi$(eV) & log $gf$ & -25 6606 & 46040 & 49841 & 
82765 & 84734 & 85205 & 100012 \\
\hline
Fe\,{\sc i} &   5242.49  &  3.63  &  -0.970 &118 &125 & 113  & 127 & 127& 137& 134\\
 &   5253.03  &  2.28  &  -3.790 &--- &--- & ---  &  60 &--- &  57&  72\\
 &   5288.52  &  3.69  &  -1.510 & 93 &--- & ---  & --- & 99 & 118& ---\\
 &   5315.05  &  4.37  &  -1.400 & 69 &--- & ---  & --- &--- &  93& ---\\ 
 &   5321.11  &  4.43  &  -1.190 &--- & 92 &  67  & --- & 76 &  82&  72\\
 &   5322.04  &  2.28  &  -2.840 &111 &125 & 109  & 110 &114 & 133& ---\\
 &   5373.71  &  4.47  &  -0.710 & 89 &--- &  --- & --- &  92& 102&  96\\
 &   5389.48  &  4.42  &  -0.250 &104 & 112& ---  &---  & 112& ---&--- \\
 &   5417.03  &  4.42  &  -1.530 & 57 & ---&  62  &  68 &  68&  81&  71\\
 &   5441.34  &  4.31  &  -1.580 & 57 & ---& ---  &  73 &  70&  70&  69\\
 &   5445.04  &  4.39  &   0.040 &136 & ---&  --- & --- &--- & ---& ---\\
 &   5522.45  &  4.21  &  -1.400 & 69 & ---& ---  & --- &--- & ---& ---\\
 &   5560.21  &  4.43  &  -1.040 & 79 &  81&  71  & --- &  78& ---&  84\\
 &   5567.39  &  2.61  &  -2.560 &116 & ---& ---  &---  &--- & ---&--- \\
 &   5624.02  &  4.39  &  -1.330 &--- &--- & ---  & --- &--- & 101& 86 \\
 &   5633.95  &  4.99  &  -0.120 & 86 & ---&  --- &  95 & ---& 108& 101\\
 &   5635.82  &  4.26  &  -1.740 & 40 &  65&  57  &  64 &  62&  58&  66\\
 &   5638.26  &  4.22  &  -0.720 & 97 & ---&  102 & 114 & 111& 115& 119\\
 &   5691.50  &  4.30  &  -1.370 & 69 & ---&  --- & --- &--- &  90& ---\\
 &   5705.47  &  4.30  &  -1.360 & 52 &  72&  65  &  75 &  67&  73&  77\\
 &   5717.83  &  4.28  &  -0.979 &--- & 107&  95  & --- & ---& ---& ---\\
 &   5731.76  &  4.26  &  -1.150 & 87 & 106&  91  &  93 &  97& 105&  99\\
 &   5806.73  &  4.61  &  -0.900 & 74 &  87&  82  &  85 &  86&  89&  93\\
 &   5814.81  &  4.28  &  -1.820 & 38 &  56&  45  &  50 &  54&  54&  56\\
 &   5852.22  &  4.55  &  -1.180 & 65 & ---&  69  &  76 & 78 &  86&  88\\
 &   5855.09  &  4.61  &  -1.520 &--- & ---& ---  &  49 &--- & ---& ---\\
 &   5856.10  &  4.29  &  -1.550 &--- & ---& ---  &  62 &--- & ---& ---\\
 &   5862.35  &  4.55  &  -0.390 &--- & ---& ---  & 112 &--- & ---& ---\\
 &   5883.82  &  3.96  &  -1.210 & 95 & ---& ---  & 112 & 101& 116& ---\\
 &   5916.25  &  2.45  &  -2.990 & 97 &--- & ---  & --- &--- & ---& ---\\
 &   5934.65  &  3.93  &  -1.020 &110 & ---& 105  & 107 &112 & 123& 119\\
 &   6024.06  &  4.55  &  -0.060 &133 & ---& 136  & --- &--- & ---&--- \\
 &   6027.05  &  4.08  &  -1.090 & 91 & 106&  87  & 100 & ---& 110& 103\\
 &   6054.07  &  4.37  &  -2.310 &--- &--- &  --- & --- &--- &  31& ---\\
 &   6056.01  &  4.73  &  -0.400 & 91 &  97& ---  &  99 &  96& 100& 103\\
 &   6078.50  &  4.80  &  -0.340 &--- & ---& ---  & 105 &--- & ---& ---\\
 &   6079.01  &  4.65  &  -0.970 & 68 & ---&  70  &  75 &  76&  79&  82\\
 &   6082.71  &  2.22  &  -3.580 &--- &--- &  80  & --- &--- & ---& ---\\
 &   6093.64  &  4.61  &  -1.350 &--- & ---&  54  &  54 & 66 &  63&  65\\
 &   6094.37  &  4.65  &  -1.940 &--- & ---&  39  & --- &--- & ---& ---\\
 &   6096.66  &  3.98  &  -1.780 &--- &  75&  63  &  70 &  68&  74&  75\\
 &   6120.25  &  0.91  &  -5.950 & 21 &  64&  33  & --- &  32&  40&  51\\
 &   6151.62  &  2.18  &  -3.290 & 87 & 117&  95  & 105 & 100& 111& 108\\
 &   6157.73  &  4.08  &  -1.110 &105 & ---& ---  & --- &--- & 123& 118\\
 &   6165.36  &  4.14  &  -1.470 & 68 &  80&  73  &  78 &  82&  88&  83\\
 &   6173.34  &  2.22  &  -2.880 &115 & ---& 114  & 130 &--- & ---& ---\\
 &   6180.20  &  2.73  &  -2.650 &--- &--- &  98  & --- &--- & ---&--- \\
 &   6187.99  &  3.94  &  -1.570 & 73 &  98&  83  &  91 &  88&  85&  92\\
 &   6200.31  &  2.60  &  -2.440 &112 & ---& 118  & 123 &126 & ---& 133\\
 &   6213.43  &  2.22  &  -2.480 &126 & ---&  --- & --- &--- & ---&--- \\
 &   6253.83  &  4.73  &  -1.660 &--- &--- &  39  & --- &--- & ---& ---\\
 &   6315.81  &  4.07  &  -1.710 &--- &--- &  72  & --- &--- & ---& ---\\
 &   6322.69  &  2.59  &  -2.430 &117 & ---& 124  & 132 &--- & ---& 135\\
 &   6380.74  &  4.19  &  -1.320 &--- &--- &  89  & --- & 98 & 102& 104\\
 &   6419.95  &  4.73  &  -0.090 &--- &--- & 111  & --- &--- & ---& ---\\
 &   6436.41  &  4.19  &  -2.460 & 27 & ---&  39  &  37 & ---& ---&  41\\
 &   6469.19  &  4.83  &  -0.620 & 93 & ---&  97  & --- &100 & ---& ---\\
 &   6496.47  &  4.79  &  -0.570 &--- &--- &  89  & --- &--- & ---& ---\\
 &   6498.95  &  0.96  &  -4.630 &--- & ---& ---  & 121 &--- & ---& ---\\
 &   6518.37  &  2.83  &  -2.300 &--- &--- & ---  & 130 & ---& ---& ---\\
 &   6533.93  &  4.56  &  -1.460 &--- &--- &  57  & --- &--- & ---& ---\\
 &   6574.23  &  0.99  &  -5.020 & 73 & ---& ---  &---  & 98 & ---&106 \\
 &   6591.31  &  4.59  &  -2.070 & 38 &--- &  26  & --- &--- & ---& ---\\
 &   6597.56  &  4.79  &  -0.920 & 63 &  72&  59  &  71 &  65&  74&  76\\
 &   6608.03  &  2.28  &  -4.030 & 49 & ---&  55  &  64 & 61 & ---&  69\\
 &   6609.11  &  2.56  &  -2.690 &111 &--- & 123  & --- &--- & ---& ---\\
 &   6627.55  &  4.55  &  -1.680 &--- &--- &  60  & --- &--- & ---& ---\\
 &   6633.75  &  4.56  &  -0.800 &--- &--- &  88  & --- &--- & ---& ---\\
 &   6646.93  &  2.61  &  -3.990 & 42 & ---& ---  & --- &--- & ---& ---\\
 &   6653.85  &  4.14  &  -2.520 & 22 & ---&  25  &  35 & ---& ---&  33\\
 &   6699.14  &  4.59  &  -2.190 &--- &  41&  30  &  51 &  29&  30&  30\\
 &   6703.57  &  2.76  &  -3.160 & 76 & ---&  --- & --- &--- & ---& ---\\
 &   6704.48  &  4.22  &  -2.660 &--- &  25&  20  &  27 &  20& ---&  19\\
 &   6713.74  &  4.79  &  -1.600 & 38 &  46&  44  &  47 &  43& ---&  52\\
 &   6725.36  &  4.10  &  -2.300 &--- &--- &  39  & --- &--- &  58& ---\\
 &   6726.67  &  4.61  &  -1.000 &--- &--- &  73  & --- &--- & ---& ---\\
 &   6733.15  &  4.64  &  -1.580 &--- &--- &  53  & --- &--- & ---& ---\\
 &   6739.52  &  1.56  &  -4.950 & 40 &  74&  47  &  53 &  54&  43&  61\\
 &   6745.96  &  4.07  &  -2.770 &--- & 21 &  17  & --- & ---& ---&  25\\
 &   6750.15  &  2.42  &  -2.620 &125 & ---& ---  & --- &--- & ---& ---\\
 &   6752.71  &  4.64  &  -1.200 & 73 & ---& ---  & --- &--- &  93& ---\\
 &   6793.26  &  4.07  &  -2.470 &--- &--- &  --- &  40 &  26& ---&  40\\
 &   6806.85  &  2.73  &  -3.210 & 73 & ---& ---  &  82 & 86 & ---& 96 \\
 &   6810.26  &  4.61  &  -0.990 & 73 &  84&  74  &  79 &  82&  79&  86\\
 &   6820.37  &  4.64  &  -1.170 & 70 &  82&  74  &  77 & ---&  75&  85\\
 &   6851.64  &  1.61  &  -5.320 & 24 & ---& ---  &---  &--- & ---&--- \\
 &   6858.15  &  4.61  &  -0.930 & 80 & ---&  76  & --- &  83&  95&  81\\
Fe\,{\sc ii}  &  4993.35  &  2.81  &  -3.670 & 70 &  70&  67 & --- & 74 &  80&  70\\
  &  5132.66  &  2.81  &  -4.000 &--- &--- &  53 &  65 & 63 &  76&  51\\
  &  5234.62  &  3.22  &  -2.240 &120 &  98& 104 & 120 & 114& 132& 110\\
  &  5264.81  &  3.23  &  -3.190 &--- & ---& --- &  73 & ---& ---& ---\\
  &  5325.56  &  3.22  &  -3.170 & 77 & ---&  61 &  70 & ---&  92&  66\\
  &  5414.05  &  3.22  &  -3.620 & 61 &  58&  49 &  61 & 60 &  73&  49\\
  &  5425.25  &  3.20  &  -3.210 & 85 &  64&  62 &  81 &  76&  89&  69\\
  &  5991.37  &  3.15  &  -3.560 & 66 &  56&  52 &  64 &  62&  80&  64\\
  &  6084.10  &  3.20  &  -3.800 & 54 & ---& --- &  48 & 55 &  69&  50\\
  &  6149.25  &  3.89  &  -2.720 & 60 &  51&  55 &  60 &  61&  76&  65\\
  &  6247.55  &  3.89  &  -2.340 &--- &  61&  71 & --- &  80& 101& ---\\
  &  6416.92  &  3.89  &  -2.680 & 69 &  56&  57 &  64 &  62&  86&  66\\
  &  6432.68  &  2.89  &  -3.580 & 76 &  65&  68 &  85 &  75&  90&  74\\\hline
\end{longtable}
}

\longtab{2}{
\begin{longtable}{ccccccccc}
\caption{Observed Fe\,{\sc i} and Fe\,{\sc ii} lines.}
\\\hline\hline
   &  &  &  & \multicolumn{5}{c}{Equivalent Widths (m\AA)}\\
\cline{5 - 9}
   &  &  &  & HD & HD & HD & HD & HD \\
Element &  $\lambda$\,(\AA) &  $\chi$(eV) & log $gf$ & 101079 & 130386 & 139660 & 198590 & 212209\\
\hline
Fe\,{\sc i}   &  5242.49  &  3.63  &  -0.970 & 115 &  119 & 124 & 127 & 132 \\
 &   5253.03  &  2.28  &  -3.790 &  63 & ---  &   74& --- & --- \\
 &   5288.52  &  3.69  &  -1.510 &  97 & 100  & 114 & --- & --- \\
 &   5321.11  &  4.43  &  -1.190 &  67 &  76  &  78 &  74 &  84 \\
 &   5322.04  &  2.28  &  -2.840 & 107 & 119  & 119 & 114 & 131 \\
 &   5373.71  &  4.47  &  -0.710 &  86 &  87  &  91 &  94 &  95 \\
 &   5389.48  &  4.42  &  -0.250 & 108 &---   &  ---& 112 & --- \\
 &   5417.03  &  4.42  &  -1.530 &  64 & ---  &  71 &  72 &  77 \\
 &   5441.34  &  4.31  &  -1.580 &  62 &  68  &  70 &  67 &  72 \\
 &   5445.04  &  4.39  &   0.041 & 134 &---   &  ---& --- & --- \\
 &   5522.45  &  4.21  &  -1.400 & --- & ---  &  80 & --- &  87 \\
 &   5531.98  &  4.91  &  -1.460 &  42 &---   &  ---& --- & --- \\
 &   5560.21  &  4.43  &  -1.040 &  74 & 77   &   78&  83 &  83 \\
 &   5584.77  &  3.57  &  -2.170 &  76 &---   &  ---& --- & --- \\
 &   5624.02  &  4.39  &  -1.330 &  83 &---   &   92&  91 & --- \\
 &   5633.95  &  4.99  &  -0.120 &  92 & 102  & --- & --- &  98 \\
 &   5635.82  &  4.26  &  -1.740 &  60 &  64  &  67 &  60 &  69 \\
 &   5638.26  &  4.22  &  -0.720 & 111 & 115  & 114 & 111 & 119 \\
 &   5705.47  &  4.30  &  -1.360 &  67 &  72  &  75 &  70 &  81 \\
 &   5717.83  &  4.28  &  -0.979 & --- & ---  & 109 & --- & --- \\
 &   5731.76  &  4.26  &  -1.150 &  92 & 100  &  95 &  92 & 103 \\
 &   5806.73  &  4.61  &  -0.900 &  82 &  89  &  90 &  82 &  92 \\
 &   5814.81  &  4.28  &  -1.820 &  47 &  53  &  58 &  48 &  61 \\
 &   5852.22  &  4.55  &  -1.180 &  74 & ---  &  89 &  77 & --- \\
 &   5883.82  &  3.96  &  -1.210 & 105 & 105  & 104 & 104 & 103 \\
 &   5916.25  &  2.45  &  -2.990 & 106 & ---  & --- & --- & --- \\
 &   5934.65  &  3.93  &  -1.020 & 108 & 117  & 116 & 116 & 125 \\
 &   6027.05  &  4.08  &  -1.090 &  96 & 103  & --- &  98 & 106 \\
 &   6056.01  &  4.73  &  -0.400 &  95 & ---  & 100 & 102 & 104 \\
 &   6079.01  &  4.65  &  -0.970 &  68 &  70  &  84 &  80 & --- \\
 &   6082.71  &  2.22  &  -3.580 & --- & ---  & --- &  93 & --- \\
 &   6093.64  &  4.61  &  -1.350 &  56 &  56  &  61 &  63 &  70 \\
 &   6096.66  &  3.98  &  -1.780 &  68 &  75  &  75 &  69 &  78 \\
 &   6120.25  &  0.91  &  -5.950 & --- &  53  & --- &  39 &  67 \\
 &   6151.62  &  2.18  &  -3.290 &  97 & 108  & 109 & 102 & 119 \\
 &   6157.73  &  4.08  &  -1.110 & 105 & ---  & 122 & 117 & --- \\
 &   6165.36  &  4.14  &  -1.470 &  74 &  77  &  81 &  80 &  85 \\
 &   6173.34  &  2.22  &  -2.880 & 120 & ---  & 139 & 132 & --- \\
 &   6187.99  &  3.94  &  -1.570 &  80 &  90  &  91 &  89 & 105 \\
 &   6200.31  &  2.60  &  -2.440 & 120 & ---  & 133 & 123 & --- \\
 &   6213.43  &  2.22  &  -2.480 & 136 &---   &  ---& --- & --- \\
 &   6322.69  &  2.59  &  -2.430 & 124 & ---  & --- & --- & --- \\
 &   6380.74  &  4.19  &  -1.320 &  92 & 103  & --- & 100 & --- \\
 &   6436.41  &  4.19  &  -2.460 &  36 &---   &   47&  40 &  53 \\
 &   6469.19  &  4.83  &  -0.620 &  96 &---   &  ---& 103 & --- \\
 &   6518.37  &  2.83  &  -2.300 & --- &109   &  ---& --- & 122 \\
 &   6574.23  &  0.99  &  -5.020 & --- &107   &  ---&  95 & 130 \\
 &   6591.31  &  4.59  &  -2.070 &  28 & 36   &  31 & --- & --- \\
 &   6597.56  &  4.79  &  -0.920 &  69 &  74  &  73 &  74 &  82 \\
 &   6608.03  &  2.28  &  -4.030 & --- &  68  & --- & --- & --- \\
 &   6609.11  &  2.56  &  -2.690 & 125 &---   &  ---& --- & --- \\
 &   6646.93  &  2.61  &  -3.990 & --- &---   &  ---&  58 & --- \\
 &   6653.85  &  4.14  &  -2.520 &  27 & 34   &   41&  29 &  39 \\
 &   6699.14  &  4.59  &  -2.190 &  25 &  38  &  37 &  29 &  43 \\
 &   6703.57  &  2.76  &  -3.160 & --- &---   &  ---& 100 & --- \\
 &   6704.48  &  4.22  &  -2.660 &  19 &  27  &  22 &  17 & --- \\
 &   6713.74  &  4.79  &  -1.600 &  41 & ---  &  52 &  47 &  56 \\
 &   6739.52  &  1.56  &  -4.950 & --- &  64  & --- &  57 &  76 \\
 &   6745.96  &  4.07  &  -2.770 & --- & 22   &  23 &  21 &  35 \\
 &   6752.71  &  4.64  &  -1.200 & --- & ---  & --- &  92 & --- \\
 &   6793.26  &  4.07  &  -2.470 &  33 &---   &  46 & --- &  51 \\
 &   6806.85  &  2.73  &  -3.210 &  93 &  91  & --- &  92 & --- \\
 &   6810.26  &  4.61  &  -0.990 &  75 &  85  &  88 &  83 &  92 \\
 &   6820.37  &  4.64  &  -1.170 &  72 &  80  &  90 &  82 &  89 \\
 &   6858.15  &  4.61  &  -0.930 &  81 &  93  &  89 &  86 & --- \\
Fe\,{\sc ii}  &  4993.35  &  2.81  &  -3.670 &  64 &  65&  69 & 65 &  59 \\
  &  5132.66  &  2.81  &  -4.000 &  48 & ---&  59 &  66 & --- \\
  &  5197.56  &  3.23  &  -2.250 & --- &--- &  117& --- & --- \\
  &  5234.62  &  3.22  &  -2.240 & 112 &110 &  107& 119 & 103 \\
  &  5284.10  &  2.89  &  -3.010 & --- & ---& --- & --- &  77 \\
  &  5325.56  &  3.22  &  -3.170 &  63 &  61& --- &  73 &  66 \\
  &  5414.05  &  3.22  &  -3.620 &  51 &  41&  55 &  64 &  53 \\
  &  5425.25  &  3.20  &  -3.210 &  66 &  57&  61 &  76 & --- \\
  &  5991.37  &  3.15  &  -3.560 &  60 &  55&  61 &  68 &  55 \\
  &  6084.10  &  3.20  &  -3.800 &  46 &  43&  47 &  60 & --- \\
  &  6149.25  &  3.89  &  -2.720 &  53 &  48&  58 &  60 &  47 \\
  &  6247.55  &  3.89  &  -2.340 & --- &--- &  69 & --- &  59 \\
  &  6416.92  &  3.89  &  -2.680 &  63 &  55&  59 &  65 &  57 \\
  &  6432.68  &  2.89  &  -3.580 &  69 &  63&  65 &  82 &  59 \\
\hline
\end{longtable}
}

\longtab{4}{
\begin{longtable}{ccccccccccc}
\caption{Other lines studied.}\\
\hline\hline
    & & & & &\multicolumn{6}{c}{Equivalent widths (m\AA)} \\
\cline{6-11}
                 &          &              &   &     &  CD & HD &  HD  & HD & HD & HD \\
$\lambda$\,(\AA) & Species  & $\chi$\,(eV) & log$gf$ & Ref & -25 6606 & 46040 & 49841  & 82765 & 84734 & 85205\\\hline

5682.65 & Na\,{\sc i} & 2.10 & -0.700 &  PS & 140 & --- & --- & --- & 148 &--- \\
5688.22 & Na\,{\sc i} & 2.10 & -0.400 &  PS & --- & --- & 151 & --- & 152 &--- \\
6154.22 & Na\,{\sc i} & 2.10 & -1.510 & R03 &  81 &  93 &  75 &  90 &  92 & 98 \\
6160.75 & Na\,{\sc i} & 2.10 & -1.210 & R03 &  89 & 123 & 111 & 107 & 100 &116 \\

4730.04 & Mg\,{\sc i} & 4.34 & -2.390 & R03 & --- & --- & --- & --- & 104 &--- \\
5711.10 & Mg\,{\sc i} & 4.34 & -1.750 & R99 & 128 & 143 & 126 & 139 & 138 &146 \\
6318.71 & Mg\,{\sc i} & 5.11 & -1.940 & Ca07&  65 &  84 &  64 &  84 & --- & 77 \\
6319.24 & Mg\,{\sc i} & 5.11 & -2.160 & Ca07& --- &  51 &  52 &  61 &  44 &--- \\
6319.49 & Mg\,{\sc i} & 5.11 & -2.670 & Ca07& --- &  30 &  15 &  39 & --- &--- \\
6965.41 & Mg\,{\sc i} & 5.75 & -1.720 & MR94&  41 &  71 &  64 &  71 &  75 & 69 \\
7387.70 & Mg\,{\sc i} & 5.75 & -0.870 & MR94&  81 & --- & --- & 118 & --- &119 \\
8712.69 & Mg\,{\sc i} & 5.93 & -1.260 & E93 & --- & --- &  71 & 103 & --- & 87 \\
8717.83 & Mg\,{\sc i} & 5.91 & -0.970 & WSM & 108 & --- &  88 & 104 & 136 &117 \\
8736.04 & Mg\,{\sc i} & 5.94 & -0.340 & WSM & 146 & 139 & 137 & 156 & --- &--- \\

6696.03 & Al\,{\sc i} & 3.14 & -1.481 & MR94& --- &  79 &  59 &  77 &  59 &--- \\
6698.67 & Al\,{\sc i} & 3.14 & -1.630 & R03 &  38 &  78 &  54 &  56 &  51 &--- \\
7835.32 & Al\,{\sc i} & 4.04 & -0.580 & R03 &  64 & --- &  62 &  80 &  64 & 91 \\
7836.13 & Al\,{\sc i} & 4.02 & -0.400 & R03 &  69 &  91 &  79 &  95 &  77 & 95 \\
8772.88 & Al\,{\sc i} & 4.02 & -0.250 & R03 & 100 & 108 & 101 & 102 & 125 &--- \\
8773.91 & Al\,{\sc i} & 4.02 & -0.070 & R03 & 123 & --- & 103 & 135 & --- &--- \\

5793.08 & Si\,{\sc i} & 4.93 & -2.060 & R03 &  69 & --- & --- &  83 &  78 & 90 \\
6125.03 & Si\,{\sc i} & 5.61 & -1.540 & E93 &  51 & --- &  52 &  57 &  58 & 59 \\
6131.58 & Si\,{\sc i} & 5.62 & -1.685 & E93 & --- &  37 &  38 &  54 & --- & 56 \\
6145.02 & Si\,{\sc i} & 5.61 & -1.430 & E93 &  58 &  47 &  52 &  61 &  57 & 68 \\
6155.14 & Si\,{\sc i} & 5.62 & -0.770 & E93 & 104 &  97 &  97 & 115 & 114 &112 \\
7760.64 & Si\,{\sc i} & 6.20 & -1.280 & E93 &  42 &  18 &  25 &  36 &  61 & 39 \\
7800.00 & Si\,{\sc i} & 6.18 & -0.720 & E93 & --- & --- &  59 &  83 & 100 &100 \\
8728.01 & Si\,{\sc i} & 6.18 & -0.360 & E93 & 105 & --- &  85 & 106 & --- &105 \\
8742.45 & Si\,{\sc i} & 5.87 & -0.510 & E93 & 114 &  88 & 111 & 124 & 121 &137 \\

6161.30 & Ca\,{\sc i} & 2.52 & -1.270 & E93   &  96 & --- & --- & 107 & --- &113 \\
6166.44 & Ca\,{\sc i} & 2.52 & -1.140 & R03   &  99 & 116 & 105 & 110 & 115 &110 \\
6169.04 & Ca\,{\sc i} & 2.52 & -0.800 & R03   & 121 & 153 & 130 & 138 & 133 &149 \\
6169.56 & Ca\,{\sc i} & 2.53 & -0.480 & DS91  & 146 & --- & 146 & 154 & 143 &--- \\
6455.60 & Ca\,{\sc i} & 2.51 & -1.290 & R03   &  89 & 125 & 104 & 108 & 103 &114 \\
6471.66 & Ca\,{\sc i} & 2.51 & -0.690 & \,S86 & 125 & 146 & 128 & 136 & 132 &--- \\

4758.12 & Ti\,{\sc i} & 2.25 &  0.425 & MFK      &  75 & --- &  75 & --- & --- &--- \\
4759.28 & Ti\,{\sc i} & 2.25 &  0.514 & MFK      & --- & --- &  79 &  87 & --- &--- \\
4820.41 & Ti\,{\sc i} & 1.50 & -0.439 & MFK      & --- & --- & --- & --- & 109 &--- \\
4997.10 & Ti\,{\sc i} & 0.00 & -2.118 & MFK      & --- & --- &  85 &  92 &  87 &--- \\
5016.17 & Ti\,{\sc i} & 0.85 & -0.574 & MFK      & 114 & --- & --- & 129 & 127 &--- \\
5022.87 & Ti\,{\sc i} & 0.83 & -0.434 & MFK      & --- & --- & 125 & 128 & 128 &--- \\
5039.96 & Ti\,{\sc i} & 0.02 & -1.130 & MFK      & --- & --- & 134 & 144 & --- &--- \\
5043.59 & Ti\,{\sc i} & 0.84 & -1.733 & MFK      & --- & --- &  75 &  82 & --- &--- \\
5062.10 & Ti\,{\sc i} & 2.16 & -0.464 & MFK      & --- & --- &  48 & --- &  55 & 42 \\
5113.45 & Ti\,{\sc i} & 1.44 & -0.880 & E93      &  60 &  91 &  63 &  72 &  61 &--- \\
5145.47 & Ti\,{\sc i} & 1.46 & -0.574 & MFK      &  72 & 112 &  83 &  86 &  85 & 86 \\
5152.19 & Ti\,{\sc i} & 0.02 & -2.024 & MFK      & --- & --- &  86 & --- &  92 &--- \\
5219.71 & Ti\,{\sc i} & 0.02 & -2.292 & MFK      & --- & --- &  95 &  99 &  93 & 91 \\
5223.63 & Ti\,{\sc i} & 2.09 & -0.559 & MFK      &  43 &  74 &  53 &  54 &  44 & 41 \\
5295.78 & Ti\,{\sc i} & 1.05 & -1.633 & MFK      &  40 &  79 &  48 &  54 &  51 & 55 \\
5490.16 & Ti\,{\sc i} & 1.46 & -0.937 & MFK      & --- & --- & --- & --- &  71 & 66 \\
5662.16 & Ti\,{\sc i} & 2.32 & -0.109 & MFK      & --- &  86 & --- &  77 &  74 & 74 \\
5689.48 & Ti\,{\sc i} & 2.30 & -0.469 & MFK      &  39 &  66 &  40 &  58 &  41 & 51 \\
5866.46 & Ti\,{\sc i} & 1.07 & -0.871 & E93      &  97 & --- & 105 & 119 & 119 &--- \\
5922.12 & Ti\,{\sc i} & 1.05 & -1.465 & MFK      & --- &  95 &  66 &  70 &  82 &--- \\
5978.55 & Ti\,{\sc i} & 1.87 & -0.496 & MFK      &  54 & 102 &  64 &  60 &  70 & 61 \\
6091.18 & Ti\,{\sc i} & 2.27 & -0.370 & R03      &  38 & --- &  51 &  59 &  51 & 53 \\
6126.22 & Ti\,{\sc i} & 1.05 & -1.370 & R03      &  63 & --- &  72 &  80 &  79 & 76 \\
6258.11 & Ti\,{\sc i} & 1.44 & -0.355 & MFK      &  97 & --- & 102 & 113 & 111 &118 \\
6261.10 & Ti\,{\sc i} & 1.43 & -0.480 & B86      &  94 & --- & 112 & 121 & 108 &119 \\
6554.24 & Ti\,{\sc i} & 1.44 & -1.219 & MFK      &  43 & --- &  64 &  68 &  59 & 51 \\

4789.34 & Cr\,{\sc i} & 2.54 & -0.365 & MFK      & --- & --- & 120 & --- & --- &--- \\
4801.03 & Cr\,{\sc i} & 3.12 & -0.130 & MFK      &  84 & --- &  83 & 100 &  96 &108 \\
4814.26 & Cr\,{\sc i} & 3.09 & -1.211 & MFK      & --- & --- &  50 & --- &  42 &--- \\
4836.85 & Cr\,{\sc i} & 3.10 & -1.137 & MFK      &  43 & --- & --- & --- &  63 &--- \\
4936.34 & Cr\,{\sc i} & 3.11 & -0.220 & MFK      & --- & --- & --- & --- & --- & 86 \\
4964.93 & Cr\,{\sc i} & 0.94 & -2.526 & MFK      & --- & 107 & --- & --- & --- &--- \\
5193.50 & Cr\,{\sc i} & 3.42 & -0.720 & MFK      & --- & --- &  36 &  41 &  31 & 30 \\
5200.18 & Cr\,{\sc i} & 3.38 & -0.650 & MFK      & --- & --- &  63 &  59 &  62 &--- \\
5214.13 & Cr\,{\sc i} & 3.37 & -0.740 & MFK      & --- & --- &  35 &  39 &  36 & 39 \\
5238.96 & Cr\,{\sc i} & 2.71 & -1.305 & MFK      &  35 &  62 &  46 & --- &  42 & 52 \\
5247.57 & Cr\,{\sc i} & 0.96 & -1.630 & MFK      & --- & --- & --- & 146 & 141 &--- \\
5272.00 & Cr\,{\sc i} & 3.45 & -0.421 & MFK      & --- &  77 &  47 &  49 &  47 &--- \\
5296.70 & Cr\,{\sc i} & 0.98 & -1.390 & GS       & 142 & --- & 144 & 157 & --- &--- \\
5300.75 & Cr\,{\sc i} & 0.98 & -2.130 & GS       & --- & 134 & 107 & 116 & 114 &123 \\
5304.18 & Cr\,{\sc i} & 3.46 & -0.692 & MFK      & --- & --- &  34 &  49 &  39 & 30 \\
5318.77 & Cr\,{\sc i} & 3.44 & -0.688 & MFK      & --- & --- & --- &  43 &  38 &--- \\
5348.33 & Cr\,{\sc i} & 1.00 & -1.290 & GS       & 136 & --- & 150 & --- & --- &--- \\
5628.65 & Cr\,{\sc i} & 3.42 & -0.772 & MFK      & --- & --- &  33 &  37 &  23 &--- \\
5702.32 & Cr\,{\sc i} & 3.45 & -0.666 & MFK      &  52 & --- &  70 &  63 &  60 &--- \\
5783.07 & Cr\,{\sc i} & 3.32 & -0.500 & MFK      &  50 &  73 &  60 &  64 &  60 & 63 \\
5783.87 & Cr\,{\sc i} & 3.32 & -0.290 & GS       &  77 & --- &  86 &  91 &  97 &--- \\
5784.97 & Cr\,{\sc i} & 3.32 & -0.379 & MFK      & --- &  80 & --- &  63 &  66 &--- \\
5787.93 & Cr\,{\sc i} & 3.32 & -0.080 & GS       &  66 &  98 &  78 &  84 &  82 & 85 \\
6330.10 & Cr\,{\sc i} & 0.94 & -2.920 & R03      &  68 & 121 &  83 &  83 &  87 & 89 \\

4904.42 & Ni\,{\sc i} & 3.54 & -0.170 & MFK      & --- & --- & 115 & --- & --- &--- \\
4913.98 & Ni\,{\sc i} & 3.74 & -0.620 & MFK      &  77 & --- &  82 &  93 &  83 &100 \\
4935.83 & Ni\,{\sc i} & 3.94 & -0.360 & MFK      &  82 &  87 &  81 &  93 &  91 & 98 \\
4953.21 & Ni\,{\sc i} & 3.74 & -0.660 & MFK      &  86 & --- &  87 &  96 &  94 & 95 \\
4967.52 & Ni\,{\sc i} & 3.80 & -1.570 & MFK      & --- & --- & --- &  39 &  32 &--- \\
4995.66 & Ni\,{\sc i} & 3.63 & -1.580 & MFK      & --- &  51 &  38 & --- &  45 &--- \\
5010.94 & Ni\,{\sc i} & 3.63 & -0.870 & MFK      &  74 &  80 & --- &  83 &  78 & 86 \\
5084.11 & Ni\,{\sc i} & 3.68 & -0.180 & E93      & 107 & --- & 102 & 124 & 107 &138 \\
5094.42 & Ni\,{\sc i} & 3.83 & -1.080 & MFK      &  49 &  59 &  56 &  63 &  55 &--- \\
5115.40 & Ni\,{\sc i} & 3.83 & -0.280 & R03      & 103 & --- &  97 & --- & 111 &125 \\
5157.98 & Ni\,{\sc i} & 3.61 & -1.590 & MFK      & --- & --- &  44 &  50 &  46 &--- \\
5197.17 & Ni\,{\sc i} & 3.90 & -1.190 & MFK      & --- & --- &  59 &  70 &  61 & 62 \\
5578.73 & Ni\,{\sc i} & 1.68 & -2.640 & MFK      & 107 & --- & 104 & 118 & 113 &121 \\
5589.37 & Ni\,{\sc i} & 3.90 & -1.140 & MFK      & --- &  56 &  46 &  53 &  48 & 61 \\
5593.75 & Ni\,{\sc i} & 3.90 & -0.840 & MFK      &  70 &  73 &  74 &  78 &  73 & 83 \\
5643.09 & Ni\,{\sc i} & 4.17 & -1.250 & MFK      & --- &  38 &  26 &  38 &  30 &--- \\
5748.36 & Ni\,{\sc i} & 1.68 & -3.260 & MFK      & --- & --- &  74 &  83 &  78 &--- \\
5760.84 & Ni\,{\sc i} & 4.11 & -0.800 & MFK      &  64 & --- &  73 &  58 &  71 & 80 \\
5805.23 & Ni\,{\sc i} & 4.17 & -0.640 & MFK      &  61 &  66 &  64 &  72 &  66 & 83 \\
5996.74 & Ni\,{\sc i} & 4.24 & -1.060 & MFK      &  38 & --- &  46 &  53 &  48 & 57 \\
6053.69 & Ni\,{\sc i} & 4.24 & -1.070 & MFK      & --- &  52 & --- & --- &  54 & 58 \\
6086.29 & Ni\,{\sc i} & 4.27 & -0.510 & MFK      &  59 &  76 &  72 &  83 &  73 &--- \\
6108.12 & Ni\,{\sc i} & 1.68 & -2.440 & MFK      & 116 & --- & 117 & 120 & 122 &--- \\
6111.08 & Ni\,{\sc i} & 4.09 & -0.870 & MFK      &  55 &  67 &  61 &  64 &  61 & 72 \\
6128.98 & Ni\,{\sc i} & 1.68 & -3.320 & MFK      &  74 & --- &  76 &  84 &  84 & 94 \\
6130.14 & Ni\,{\sc i} & 4.27 & -0.960 & MFK      &  40 &  47 &  40 &  50 &  47 & 49 \\
6176.82 & Ni\,{\sc i} & 4.09 & -0.264 & R03      &  88 &  97 &  85 & 100 &  93 &114 \\
6177.25 & Ni\,{\sc i} & 1.83 & -3.510 & MFK      &  45 &  67 &  49 &  56 &  52 & 66 \\
6186.72 & Ni\,{\sc i} & 4.11 & -0.960 & MFK      &  51 &  60 &  52 &  64 &  54 & 58 \\
6204.61 & Ni\,{\sc i} & 4.09 & -1.140 & MFK      &  47 & --- &  53 &  61 &  58 & 59 \\
6223.99 & Ni\,{\sc i} & 4.11 & -0.980 & MFK      & --- & --- &  60 &  61 &  48 &--- \\
6230.10 & Ni\,{\sc i} & 4.11 & -1.260 & MFK      & --- & --- & --- &  52 &  57 & 61 \\
6322.17 & Ni\,{\sc i} & 4.15 & -1.170 & MFK      &  32 & --- &  37 &  47 &  34 &--- \\
6327.60 & Ni\,{\sc i} & 1.68 & -3.113 & MFW      &  88 & --- & --- & 109 & 103 &--- \\
6378.26 & Ni\,{\sc i} & 4.15 & -0.900 & MFK      &  56 & --- & --- &  71 &  72 &--- \\
6384.67 & Ni\,{\sc i} & 4.15 & -1.130 & MFK      &  47 & --- &  43 &  52 &  46 &--- \\
6482.80 & Ni\,{\sc i} & 1.94 & -2.630 & MFW      &  81 & --- & 100 & 108 & 101 &--- \\
6532.88 & Ni\,{\sc i} & 1.94 & -3.390 & MFK      &  67 &  69 &  50 &  64 & 124 &--- \\
6586.33 & Ni\,{\sc i} & 1.95 & -2.810 & MFW      &  85 & --- &  90 &  98 &  96 &--- \\
6598.61 & Ni\,{\sc i} & 4.24 & -0.980 & MFK      &  42 & --- &  43 &  57 &  47 &--- \\
6635.14 & Ni\,{\sc i} & 4.42 & -0.830 & MFK      &  50 & --- & --- & --- &  55 &--- \\
6643.64 & Ni\,{\sc i} & 1.68 & -2.030 & MFW      & 142 & --- & --- & --- & --- &--- \\
6767.77 & Ni\,{\sc i} & 1.83 & -2.170 & MFW      & 125 & --- & --- & --- & --- &--- \\
6772.32 & Ni\,{\sc i} & 3.66 & -0.970 & R03      &  70 &  84 &  80 & --- &  98 &--- \\
6842.04 & Ni\,{\sc i} & 3.66 & -1.477 & MFK      &  47 &  65 &  69 & --- &  58 &--- \\
7788.93 & Ni\,{\sc i} & 1.95 & -1.990 & E93      & --- & --- & 141 & --- & --- &--- \\

4883.68 & Y\,{\sc ii} & 1.08 &  0.070 & SN96  & --- & --- & 145 & --- & --- &--- \\
5087.43 & Y\,{\sc ii} & 1.08 & -0.170 & SN96  & 123 & 142 & 115 & 124 & 119 &--- \\
5200.41 & Y\,{\sc ii} & 0.99 & -0.570 & SN96  & --- & --- & 109 & 112 & 116 &--- \\
5205.72 & Y\,{\sc ii} & 1.03 & -0.340 & SN96  & --- & --- & 134 & --- & --- &--- \\
5289.81 & Y\,{\sc ii} & 1.03 & -1.850 & VWR   &  49 &  74 &  55 &  46 &  56 & 64 \\
5402.78 & Y\,{\sc ii} & 1.84 & -0.440 & R03   &  78 &  85 &  66 &  66 &  75 & 93 \\

6127.46 & Zr\,{\sc i} & 0.15 & -1.060 & H82   &  20 &  96 &  36 &  29 &  36 & 39 \\
6134.57 & Zr\,{\sc i} & 0.00 & -1.280 & B81   &  21 &  93 &  35 &  30 &  36 & 32 \\
6140.46 & Zr\,{\sc i} & 0.52 & -1.410 & S96   & --- &  43 &  12 & --- & --- &--- \\
6143.18 & Zr\,{\sc i} & 0.07 & -1.100 & B81   &  24 & 106 &  46 &  40 &  43 & 39 \\

4934.83 & La\,{\sc ii} & 1.25 & -0.920 & VWR  &  35 & --- &  28 & --- &  34 & 37 \\
5303.53 & La\,{\sc ii} & 0.32 & -1.350 & VWR  &  54 & --- &  42 &  39 &  42 & 50 \\
5880.63 & La\,{\sc ii} & 0.23 & -1.830 & R04  & --- &  87 &  36 & --- &  36 &--- \\
6320.42 & La\,{\sc ii} & 0.17 & -1.520 & VWR  &  41 & 124 &  47 &  38 &  65 & 47 \\
6390.48 & La\,{\sc ii} & 0.32 & -1.410 & S96  &  54 & 115 &  55 &  42 &  53 & 53 \\
6774.33 & La\,{\sc ii} & 0.12 & -1.709 & S96  &  52 & 119 &  44 &  39 &  47 & 50 \\

4562.37 & Ce\,{\sc ii} & 0.48 &  0.330 & S96 &  87 & 116 & --- & --- &  85 &--- \\
4628.16 & Ce\,{\sc ii} & 0.52 &  0.260 & S96 & --- & --- & --- & --- &  95 &--- \\
5117.17 & Ce\,{\sc ii} & 1.40 &  0.010 & VWR &  28 & --- &  26 & --- & --- & 38 \\
5187.46 & Ce\,{\sc ii} & 1.21 &  0.300 & VWR &  51 &  86 & --- & --- &  47 & 57 \\
5274.24 & Ce\,{\sc ii} & 1.28 &  0.389 & VWR &  58 &  80 &  53 &  49 &  57 & 72 \\
5330.58 & Ce\,{\sc ii} & 0.87 &  0.280 & VWR & --- & --- & --- &  36 & --- &--- \\
5472.30 & Ce\,{\sc ii} & 1.25 & -0.190 & VWR &  40 &  61 &  33 &  24 &  40 & 30 \\
6051.80 & Ce\,{\sc ii} & 0.23 & -1.600 & S96 &  20 &  51 &  17 &  10 &  20 &--- \\

4811.34 & Nd\,{\sc ii} & 0.06 & -1.015 & VWR &  71 & --- &  67 & --- &  69 & 87 \\
4959.12 & Nd\,{\sc ii} & 0.06 & -0.916 & VWR &  86 & --- & --- & --- & --- &105 \\
4989.95 & Nd\,{\sc ii} & 0.63 & -0.624 & VWR &  57 & --- & --- & --- & --- &--- \\
5063.72 & Nd\,{\sc ii} & 0.98 & -0.758 & VWR & --- & --- & --- & --- &  32 &--- \\
5092.80 & Nd\,{\sc ii} & 0.38 & -0.510 & E93 & --- &  91 & --- & --- & --- &--- \\
5130.59 & Nd\,{\sc ii} & 1.30 &  0.100 & SN96& --- & --- & --- &  38 & --- &--- \\
5212.36 & Nd\,{\sc ii} & 0.20 & -0.700 & E93 &  79 & --- &  83 & --- &  92 &--- \\
5234.19 & Nd\,{\sc ii} & 0.55 & -0.460 & S96 & --- & --- & --- & --- & --- & 73 \\
5311.46 & Nd\,{\sc ii} & 0.98 & -0.560 & SN96& --- & --- &  49 & --- & --- &--- \\
5319.81 & Nd\,{\sc ii} & 0.55 & -0.350 & SN96& --- & --- &  74 &  63 & --- &--- \\
5416.38 & Nd\,{\sc ii} & 0.86 & -0.980 & VWR & --- &  50 & --- & --- & --- &--- \\
5431.54 & Nd\,{\sc ii} & 1.12 & -0.457 & VWR & --- &  58 &  29 & --- &  33 &--- \\
5442.26 & Nd\,{\sc ii} & 0.68 & -0.900 & SN96&  46 & --- & --- & --- &  54 & 65 \\
5740.88 & Nd\,{\sc ii} & 1.16 & -0.560 & VWR & --- & --- &  26 &  29 &  40 & 32 \\
5842.39 & Nd\,{\sc ii} & 1.28 & -0.601 & VWR & --- &  40 & --- & --- & --- &--- \\
\hline
\\
\end{longtable}
}

\longtab{4}{
\begin{longtable}{ccccccccccc}
\caption{Other lines studied.}\\
\hline\hline
 & & & & &\multicolumn{6}{c}{Equivalent widths (m\AA)} \\
\cline{6 -11}
                 &          &              &   &   & HD &  HD  &  HD   &  HD  & HD &  HD \\
$\lambda$\,(\AA) & Species  & $\chi$\,(eV) & log$gf$ & Ref & 100012 & 101079 &  130386 & 139660 & 198590 & 212209\\
\hline
6154.22 & Na\,{\sc i} & 2.10 & -1.510 & R03 & 106 &  91& 111&  106 &  94 & 117 \\
6160.75 & Na\,{\sc i} & 2.10 & -1.210 & R03 & 120 & 106& 125&  123 & 112 & 124 \\

4730.04 & Mg\,{\sc i} & 4.34 & -2.390 & R03  & 121 & ---& 114&  --- & 113 & --- \\
5711.10 & Mg\,{\sc i} & 4.34 & -1.750 & R99  & 141 & 132& 143&  148 & 138 & 156 \\
6318.71 & Mg\,{\sc i} & 5.11 & -1.940 & Ca07 &  76 &  69&  84&   76 &  78 & --- \\
6319.24 & Mg\,{\sc i} & 5.11 & -2.160 & Ca07 &  59 & ---&  63&  --- & --- &  81 \\
6319.49 & Mg\,{\sc i} & 5.11 & -2.670 & Ca07 &  46 &  37&  41&   40 &  41 &  49 \\
6765.45 & Mg\,{\sc i} & 5.75 & -1.940 & MR94 &  35 &  40& ---&  --- &  32 &  49 \\
6965.41 & Mg\,{\sc i} & 5.75 & -1.720 & MR94 &  91 &  65&  77&   82 &  73 &  83 \\
7387.70 & Mg\,{\sc i} & 5.75 & -0.870 & MR94 &  98 &  97& 102&  107 & --- & --- \\
8712.69 & Mg\,{\sc i} & 5.93 & -1.260 & WSM  & 110 & ---&  96&  --- & --- &  89 \\
8717.83 & Mg\,{\sc i} & 5.91 & -0.970 & WSM  & 124 & 116& 114&  112 & 125 & 106 \\
8736.04 & Mg\,{\sc i} & 5.94 & -0.340 & WSM  & --- & 159& 161&  156 & --- & 165 \\

6696.03 & Al\,{\sc i} & 3.14 & -1.481 &  MR &  91 &  79&  90&   90 &  79 & 104 \\
6698.67 & Al\,{\sc i} & 3.14 & -1.630 & R03 & --- & ---&  80&   70 & --- &  90 \\
7835.32 & Al\,{\sc i} & 4.04 & -0.580 & R03 &  90 &  85&  92&   82 &  84 & 101 \\
7836.13 & Al\,{\sc i} & 4.02 & -0.400 & R03 &  94 &  94& 101&   99 &  94 & 113 \\
8772.88 & Al\,{\sc i} & 4.02 & -0.250 & R03 & 121 &--- & 110&  135 & 127 & 132 \\
8773.91 & Al\,{\sc i} & 4.02 & -0.070 & R03 & --- &140 & 152&  --- & --- & 162 \\

5793.08 & Si\,{\sc i} & 4.93 & -2.060 & R03 &  75 &  73&  80&   81 &  82 & --- \\
6125.03 & Si\,{\sc i} & 5.61 & -1.540 & E93 &  58 &  56&  70&   66 &  62 &  71 \\
6131.58 & Si\,{\sc i} & 5.62 & -1.685 & E93 &  46 &  50&  54&   55 &  59 &  47 \\
6145.02 & Si\,{\sc i} & 5.61 & -1.430 & E93 &  61 &  60&  58&   63 &  63 &  58 \\
6155.14 & Si\,{\sc i} & 5.62 & -0.770 & E93 & 110 & 100& 105&  108 & 109 & 110 \\
7760.64 & Si\,{\sc i} & 6.20 & -1.280 & E93 &  39 &  31& ---&   51 &  45 &  36 \\
7800.00 & Si\,{\sc i} & 6.18 & -0.720 & E93 &  93 &  80&  78&  --- &  95 &  86 \\
8728.01 & Si\,{\sc i} & 6.18 & -0.360 & E93 & 107 & 111&  93&  104 & 118 &  99 \\
8742.45 & Si\,{\sc i} & 5.87 & -0.510 & E93 & 102 & 114& 106&  129 & 127 & --- \\

6161.30 & Ca\,{\sc i} & 2.52 & -1.270 & E93   & --- & ---& ---&  122 & 114 & 124 \\
6166.44 & Ca\,{\sc i} & 2.52 & -1.140 & R03   & 117 & 107& 115&  113 & 109 & 122 \\
6169.04 & Ca\,{\sc i} & 2.52 & -0.800 & R03   & 140 & 131& 146&  147 & 140 & 159 \\
6169.56 & Ca\,{\sc i} & 2.53 & -0.480 & DS91  & 157 & 147& ---&  --- & --- & --- \\
6455.60 & Ca\,{\sc i} & 2.51 & -1.290 & R03   & 109 & 100& 118&  117 & 106 & 126 \\
6471.66 & Ca\,{\sc i} & 2.51 & -0.690 & \,S86 & 143 & 134& 147&  145 & 139 & 145 \\

4758.12 & Ti\,{\sc i} & 2.25 &  0.425 & MFK      &  89 & --- & --- & --- & --- &--- \\
4759.28 & Ti\,{\sc i} & 2.25 &  0.514 & MFK      & --- &  85 &  94 &  92 & --- &101 \\
4997.10 & Ti\,{\sc i} & 0.00 & -2.118 & MFK      & 106 &  87 & 103 & 102 & --- &123 \\
5009.66 & Ti\,{\sc i} & 0.02 & -2.259 & MFK      & 116 & --- & --- & --- & --- &--- \\
5016.17 & Ti\,{\sc i} & 0.85 & -0.574 & MFK      & 131 & 118 & 142 & 132 & 129 &--- \\
5022.87 & Ti\,{\sc i} & 0.83 & -0.434 & MFK      & 146 & 125 & 141 & 136 & 128 &--- \\
5039.96 & Ti\,{\sc i} & 0.02 & -1.130 & MFK      & --- & 135 & 161 & --- & 152 &--- \\
5043.59 & Ti\,{\sc i} & 0.84 & -1.733 & MFK      &  83 &  70 &  95 &  89 & --- &108 \\
5062.10 & Ti\,{\sc i} & 2.16 & -0.464 & MFK      &  59 &  52 &  71 &  64 &  46 & 81 \\
5087.06 & Ti\,{\sc i} & 1.43 & -0.840 & E93      & --- &  80 & --- & --- &  79 &--- \\
5113.45 & Ti\,{\sc i} & 1.44 & -0.880 & E93      &  85 &  69 & --- &  76 &  62 &--- \\
5145.47 & Ti\,{\sc i} & 1.46 & -0.574 & MFK      &  96 &  82 & 101 &  97 &  88 &112 \\
5147.48 & Ti\,{\sc i} & 0.00 & -2.012 & MFK      & 124 & 110 & --- & --- & --- &139 \\
5152.19 & Ti\,{\sc i} & 0.02 & -2.024 & MFK      & 121 &  91 & 110 & --- & --- &137 \\
5219.71 & Ti\,{\sc i} & 0.02 & -2.292 & MFK      & 108 &  97 & 118 & 113 &  95 &--- \\
5223.63 & Ti\,{\sc i} & 2.09 & -0.559 & MFK      &  66 &  47 &  66 &  61 &  44 & 87 \\
5295.78 & Ti\,{\sc i} & 1.05 & -1.633 & MFK      &  60 &  51 &  69 &  64 &  49 & 81 \\
5490.16 & Ti\,{\sc i} & 1.46 & -0.937 & MFK      &  83 &  63 &  90 &  83 &  61 &100 \\
5662.16 & Ti\,{\sc i} & 2.32 & -0.109 & MFK      &  81 &  70 &  87 &  84 &  73 &--- \\
5689.48 & Ti\,{\sc i} & 2.30 & -0.469 & MFK      &  54 &  48 &  67 &  62 &  39 & 72 \\
5866.46 & Ti\,{\sc i} & 1.07 & -0.871 & E93      & 129 & 116 & --- & --- & --- &--- \\
5922.12 & Ti\,{\sc i} & 1.05 & -1.465 & MFK      &  89 &  71 &  86 &  98 &  84 &106 \\
5978.55 & Ti\,{\sc i} & 1.87 & -0.496 & MFK      &  83 &  69 &  85 &  84 &  68 &100 \\
6091.18 & Ti\,{\sc i} & 2.27 & -0.370 & R03      &  61 &  52 &  76 &  68 &  57 & 83 \\
6126.22 & Ti\,{\sc i} & 1.05 & -1.370 & R03      &  90 &  80 & 100 &  93 &  77 &111 \\
6258.11 & Ti\,{\sc i} & 1.44 & -0.355 & MFK      & 119 & 105 & 123 & 113 & 111 &136 \\
6261.10 & Ti\,{\sc i} & 1.43 & -0.480 & B86      & 136 & 113 & --- & --- & 115 &--- \\
6554.24 & Ti\,{\sc i} & 1.44 & -1.219 & MFK      &  88 &  66 &  92 &  85 &  61 &115 \\

4789.34 & Cr\,{\sc i} & 2.54 & -0.365 & MFK      & --- & --- & --- & 116 & --- &--- \\
4801.03 & Cr\,{\sc i} & 3.12 & -0.130 & MFK      &  96 &  90 &  96 &  95 &  91 & 99 \\
4814.26 & Cr\,{\sc i} & 3.09 & -1.211 & MFK      & --- &  51 & --- & --- & --- &--- \\
4836.85 & Cr\,{\sc i} & 3.10 & -1.137 & MFK      &  58 & --- &  65 & --- & --- & 70 \\
4936.34 & Cr\,{\sc i} & 3.11 & -0.220 & MFK      &  88 & --- & --- & --- & --- &--- \\
4964.93 & Cr\,{\sc i} & 0.94 & -2.526 & MFK      & --- & --- & --- & --- & --- &116 \\
5193.50 & Cr\,{\sc i} & 3.42 & -0.720 & MFK      &  54 &  35 &  51 &  47 &  34 & 63 \\
5200.18 & Cr\,{\sc i} & 3.38 & -0.650 & MFK      &  69 &  62 &  66 &  67 &  59 & 81 \\
5214.13 & Cr\,{\sc i} & 3.37 & -0.740 & MFK      &  42 &  34 &  45 &  45 &  36 & 61 \\
5238.96 & Cr\,{\sc i} & 2.71 & -1.305 & MFK      &  50 &  62 &  52 & --- & --- &--- \\
5243.36 & Cr\,{\sc i} & 3.39 & -0.567 & MFK      & --- &  67 & --- &  77 & --- &--- \\
5247.57 & Cr\,{\sc i} & 0.96 & -1.630 & MFK      & --- & 136 & --- & --- & 144 &--- \\
5272.00 & Cr\,{\sc i} & 3.45 & -0.421 & MFK      & --- &  46 &  54 & --- &  45 & 71 \\
5300.75 & Cr\,{\sc i} & 0.98 & -2.130 & GS       & 125 & 112 & 126 & 123 & 113 &141 \\
5304.18 & Cr\,{\sc i} & 3.46 & -0.692 & MFK      &  45 &  39 &  49 &  50 &  39 & 55 \\
5312.86 & Cr\,{\sc i} & 3.45 & -0.562 & MFK      & --- & --- & --- & --- &  38 & 67 \\
5318.77 & Cr\,{\sc i} & 3.44 & -0.688 & MFK      &  47 &  40 &  53 &  53 &  41 & 59 \\
5340.45 & Cr\,{\sc i} & 3.44 & -0.730 & MFK      & --- &  62 & --- & --- & --- & 71 \\
5628.65 & Cr\,{\sc i} & 3.42 & -0.772 & MFK      &  40 &  33 &  42 & --- &  33 & 61 \\
5702.32 & Cr\,{\sc i} & 3.45 & -0.666 & MFK      &  52 &  63 &  67 &  67 &  56 &--- \\
5781.75 & Cr\,{\sc i} & 3.32 & -0.750 & MFK      &  66 & --- & --- & --- & --- &--- \\
5783.07 & Cr\,{\sc i} & 3.32 & -0.500 & MFK      &  69 &  59 &  68 &  71 &  63 & 74 \\
5783.87 & Cr\,{\sc i} & 3.32 & -0.290 & GS       & --- & --- & --- & --- &  97 &104 \\
5784.97 & Cr\,{\sc i} & 3.32 & -0.379 & MFK      & --- &  59 &  69 &  77 &  65 & 86 \\
5787.93 & Cr\,{\sc i} & 3.32 & -0.080 & GS       &  89 &  80 &  90 &  89 &  84 & 94 \\
6330.10 & Cr\,{\sc i} & 0.94 & -2.920 & R03      &  94 &  83 & 103 &  98 &  83 &118 \\

4904.42 & Ni\,{\sc i} & 3.54 & -0.170 & MFK      & 130 & 109 & --- & --- & --- &--- \\
4913.98 & Ni\,{\sc i} & 3.74 & -0.620 & MFK      &  99 &  86 &  88 &  91 &  89 & 98 \\
4935.83 & Ni\,{\sc i} & 3.94 & -0.360 & MFK      &  87 &  88 &  86 &  92 &  93 & 88 \\
4953.21 & Ni\,{\sc i} & 3.74 & -0.660 & MFK      & 102 &  92 & 101 &  98 & 100 &--- \\
4967.52 & Ni\,{\sc i} & 3.80 & -1.570 & MFK      &  39 &  35 &  41 &  44 &  41 & 50 \\
4995.66 & Ni\,{\sc i} & 3.63 & -1.580 & MFK      &  46 &  43 &  45 &  55 &  45 &--- \\
5003.75 & Ni\,{\sc i} & 1.68 & -3.130 & MFK      &  83 & --- & --- & --- & --- & 93 \\
5010.94 & Ni\,{\sc i} & 3.63 & -0.870 & MFK      &  83 &  73 & --- & --- & --- & 81 \\
5084.11 & Ni\,{\sc i} & 3.68 & -0.180 & E93      & --- & 107 & 112 & 111 & 114 &--- \\
5094.42 & Ni\,{\sc i} & 3.83 & -1.080 & MFK      &  58 &  55 &  61 &  58 &  59 &--- \\
5115.40 & Ni\,{\sc i} & 3.83 & -0.280 & R03      & --- & --- & --- & 109 & --- &--- \\
5157.98 & Ni\,{\sc i} & 3.61 & -1.590 & MFK      &  46 &  44 &  49 &  51 &  48 & 53 \\
5197.17 & Ni\,{\sc i} & 3.90 & -1.190 & MFK      &  71 &  62 & --- &  73 &  66 &--- \\
5578.73 & Ni\,{\sc i} & 1.68 & -2.640 & MFK      & 121 & 100 & --- & --- & --- &--- \\
5587.87 & Ni\,{\sc i} & 1.94 & -2.370 & MFK      & 107 & --- & --- & --- & --- &110 \\
5589.37 & Ni\,{\sc i} & 3.90 & -1.140 & MFK      &  53 &  50 &  55 &  55 &  53 & 61 \\
5593.75 & Ni\,{\sc i} & 3.90 & -0.840 & MFK      &  76 &  68 &  74 &  76 &  77 & 83 \\
5643.09 & Ni\,{\sc i} & 4.17 & -1.250 & MFK      &  40 &  32 &  35 &  39 &  35 &--- \\
5748.36 & Ni\,{\sc i} & 1.68 & -3.260 & MFK      &  86 &  76 &  86 &  83 &  77 & 99 \\
5760.84 & Ni\,{\sc i} & 4.11 & -0.800 & MFK      &  84 &  75 &  80 &  76 & --- &--- \\
5805.23 & Ni\,{\sc i} & 4.17 & -0.640 & MFK      &  74 &  63 &  68 &  73 &  72 & 75 \\
5847.01 & Ni\,{\sc i} & 1.68 & -3.440 & MFK      &  82 & --- & --- & --- & --- & 95 \\
5996.74 & Ni\,{\sc i} & 4.24 & -1.060 & MFK      &  57 &  43 &  55 &  57 &  53 &--- \\
6053.69 & Ni\,{\sc i} & 4.24 & -1.070 & MFK      & --- &  50 &  59 &  62 &  60 &--- \\
6086.29 & Ni\,{\sc i} & 4.27 & -0.510 & MFK      &  79 &  69 &  77 &  77 &  75 & 74 \\
6108.12 & Ni\,{\sc i} & 1.68 & -2.440 & MFK      & 132 & 118 & 128 & 129 & 118 &133 \\
6111.08 & Ni\,{\sc i} & 4.09 & -0.870 & MFK      &  74 &  62 &  67 &  66 &  69 & 69 \\
6128.98 & Ni\,{\sc i} & 1.68 & -3.320 & MFK      &  90 &  79 &  92 &  93 &  89 &104 \\
6130.14 & Ni\,{\sc i} & 4.27 & -0.960 & MFK      &  58 &  46 &  49 &  54 &  58 & 52 \\
6176.82 & Ni\,{\sc i} & 4.09 & -0.264 & R03      & 100 &  91 &  93 &  95 &  99 &103 \\
6177.25 & Ni\,{\sc i} & 1.83 & -3.510 & MFK      &  62 &  51 &  59 &  61 &  59 & 69 \\
6186.72 & Ni\,{\sc i} & 4.11 & -0.960 & MFK      &  66 &  55 &  59 &  63 &  57 & 75 \\
6204.61 & Ni\,{\sc i} & 4.09 & -1.140 & MFK      &  60 &  54 &  67 &  67 &  59 &--- \\
6223.99 & Ni\,{\sc i} & 4.11 & -0.980 & MFK      &  63 &  56 &  61 &  70 &  64 & 80 \\
6230.10 & Ni\,{\sc i} & 4.11 & -1.260 & MFK      & --- &  51 &  57 &  56 & --- &--- \\
6322.17 & Ni\,{\sc i} & 4.15 & -1.170 & MFK      &  47 &  38 &  47 &  47 &  44 &--- \\
6327.60 & Ni\,{\sc i} & 1.68 & -3.113 & MFW      & 112 &  97 & --- & --- & 106 &--- \\
6378.26 & Ni\,{\sc i} & 4.15 & -0.900 & MFK      & --- &  62 & --- &  78 &  75 &--- \\
6384.67 & Ni\,{\sc i} & 4.15 & -1.130 & MFK      & --- &  46 &  49 &  53 &  49 & 57 \\
6482.80 & Ni\,{\sc i} & 1.94 & -2.630 & MFW      & 111 &  92 & 117 & 117 & 100 &--- \\
6532.88 & Ni\,{\sc i} & 1.94 & -3.390 & MFK      & --- &  49 & --- &  60 &  58 & 81 \\
6586.33 & Ni\,{\sc i} & 1.95 & -2.810 & MFW      & 103 &  95 & 103 & 105 & 101 &123 \\
6598.61 & Ni\,{\sc i} & 4.24 & -0.980 & MFK      &  58 &  52 & --- &  55 &  55 & 69 \\
6643.64 & Ni\,{\sc i} & 1.68 & -2.030 & MFW      & --- & 149 & --- & 163 & --- &--- \\
6767.77 & Ni\,{\sc i} & 1.83 & -2.170 & MFW      & --- & 139 & --- & --- & --- &--- \\
6772.32 & Ni\,{\sc i} & 3.66 & -0.970 & R03      &  96 &  90 &  96 &  98 &  97 & 94 \\
6842.04 & Ni\,{\sc i} & 3.66 & -1.477 & MFK      &  67 &  62 & --- & --- &  69 &--- \\

4883.68 & Y\,{\sc ii} & 1.08 &  0.070 & SN96  & --- &--- & ---&  154 & --- & --- \\
5087.43 & Y\,{\sc ii} & 1.08 & -0.170 & SN96  &  93 &114 & 118&  130 & 131 & 117 \\
5200.41 & Y\,{\sc ii} & 0.99 & -0.570 & SN96  &  99 &111 & 114&  --- & 124 & 118 \\
5289.81 & Y\,{\sc ii} & 1.03 & -1.850 & VWR   &  26 &  52&  52&   49 &  58 &  54 \\
5402.78 & Y\,{\sc ii} & 1.84 & -0.440 & R03   &  50 &  68&  68&  --- &  79 &  68 \\

6127.46 & Zr\,{\sc i} & 0.15 & -1.060 & H82   &  28 &  37&  63&   52 &  37 &  68 \\
6134.57 & Zr\,{\sc i} & 0.00 & -1.280 & B81   &  27 &  36&  58&   50 &  35 &  65 \\
6140.46 & Zr\,{\sc i} & 0.52 & -1.410 & S96   & --- & ---&  20&   19 &  13 &  25 \\
6143.18 & Zr\,{\sc i} & 0.07 & -1.100 & B81   &  33 &  44&  64&   59 &  44 &  78 \\

4086.71 & La\,{\sc ii} & 0.00 & -0.160 & VWR  & --- & ---& ---&   97 &  98 & --- \\
4934.83 & La\,{\sc ii} & 1.25 & -0.920 & VWR  & --- & ---& ---&   23 & --- & --- \\
5303.53 & La\,{\sc ii} & 0.32 & -1.350 & VWR  &  33 &  40&  35&   37 &  37 &  41 \\
5880.63 & La\,{\sc ii} & 0.23 & -1.830 & R04  &  26 &  22&  38&   36 &  38 &  39 \\
6320.43 & La\,{\sc ii} & 0.17 & -1.520 & VWR  &  38 &  42&  44&   47 &  38 &  46 \\
6390.48 & La\,{\sc ii} & 0.32 & -1.410 & S96  &  36 &  47&  54&   52 &  39 & --- \\
6774.33 & La\,{\sc ii} & 0.12 & -1.709 & S96  &  30 &  38&  34&  --- &  32 &  40 \\

4073.47 & Ce\,{\sc ii} & 0.48 &  0.320 & SN96 & --- & ---& ---&   79 &  82 & --- \\
4120.84 & Ce\,{\sc ii} & 0.32 & -0.240 & S96  &  65 & ---& ---&  --- &  72 & --- \\
4486.91 & Ce\,{\sc ii} & 0.30 & -0.360 & S96  & --- &--- &  70&  --- & --- & --- \\
4562.37 & Ce\,{\sc ii} & 0.48 &  0.330 & S96  &  80 &  79&  86&   76 &  82 &  84 \\
4628.16 & Ce\,{\sc ii} & 0.52 &  0.260 & S96  &  77 & ---&  81&   76 &  90 & --- \\
5117.17 & Ce\,{\sc ii} & 1.40 &  0.010 & VWR  & --- & ---&--- &  --- &  24 & --- \\
5187.46 & Ce\,{\sc ii} & 1.21 &  0.300 & VWR  & --- &  41& ---&  --- &  40 &  40 \\
5274.24 & Ce\,{\sc ii} & 1.28 &  0.389 & VWR  &  46 &  48&  48&  --- &  47 &  44 \\
5472.30 & Ce\,{\sc ii} & 1.25 & -0.190 & VWR  &  21 &  21&  29&  --- &  22 &  20 \\
6051.80 & Ce\,{\sc ii} & 0.23 & -1.600 & S96  &  13 & 15 &  20&   18 &  15 &  21 \\

4811.34 & Nd\,{\sc ii} & 0.06 & -1.015 & VWR  &  60 &  58&  65&   57 &  58 & --- \\
4989.95 & Nd\,{\sc ii} & 0.63 & -0.624 & VWR  &  50 &  56& ---&   49 &  49 & --- \\
5063.72 & Nd\,{\sc ii} & 0.98 & -0.758 & VWR  &  16 &  19&  17&   37 &  19 & --- \\
5092.80 & Nd\,{\sc ii} & 0.38 & -0.510 & E93  & --- & ---&  49&  --- & --- &  59 \\
5212.36 & Nd\,{\sc ii} & 0.20 & -0.700 & E93  & --- & ---&--- &  --- &  78 & --- \\
5234.19 & Nd\,{\sc ii} & 0.55 & -0.460 & S96  &  58 & ---& ---&   60 &  58 & --- \\
5311.46 & Nd\,{\sc ii} & 0.98 & -0.560 & SN96 &  38 &  40& ---&   36 &  37 & --- \\
5319.81 & Nd\,{\sc ii} & 0.55 & -0.350 & SN96 &  80 &  76&  56&   66 &  72 &  61 \\
5416.38 & Nd\,{\sc ii} & 0.86 & -0.980 & VWR  & --- & 15 & ---&  --- &  16 & --- \\
5431.54 & Nd\,{\sc ii} & 1.12 & -0.457 & VWR  & --- &  27& ---&  --- &  23 & --- \\
5442.26 & Nd\,{\sc ii} & 0.68 & -0.900 & SN96 &  49 & ---& ---&  --- & --- & --- \\
5740.88 & Nd\,{\sc ii} & 1.16 & -0.560 & VWR  &  20 & 21 & ---&   28 &  20 &  33 \\
5842.39 & Nd\,{\sc ii} & 1.28 & -0.601 & VWR  & --- & 31 &  15&  --- &  11 & --- \\
\hline
%\footnote{\noident 
\footnote{References to Table 4:
\par References to Table 4:
\par B81: \cite{biemont1981}; 
\par Ca07: \cite{carretta2007};
\par DS91: \cite{drakesmith1991};
\par E93: \cite{edvardsson1993};
\par GS: \cite{grattonsneden1988};
\par H82: \cite{hannaford1982}; 
\par PS: \cite{prestonsneden2001};
\par R03: \cite{reddy2003}; 
\par R04: \cite{reyniers2004};
\par MFK: \cite{martin2002};
\par MFW: \cite{martin1988}; 
\par MR94: \cite{mcwilliam1994};
\par S86: \cite{smith1986};
\par S96: \cite{smith1996};
\par SN96: \cite{sneden1996}; 
\par VWR: \cite{vanwinckelreyniers2000};
\par WSM: \cite{wiese1969};
}
\end{longtable}
}

\end{document}